\documentclass[%
 reprint,
superscriptaddress,
 noeprint,
 amsmath,amssymb,
 aps, prl,
]{revtex4-2}

\usepackage{graphicx}%
\usepackage{dcolumn}%
\usepackage{bm}%
\usepackage{xcolor} 

\usepackage{siunitx}
\DeclareSIUnit\bohr{\text {\ensuremath {a}}_{0}}
\DeclareSIUnit\gauss{\text{G}}
\usepackage{xspace}
\usepackage{braket}
\usepackage{placeins}
\usepackage[normalem]{ulem}

\newcommand{\Cs}{$^{133}$Cs\xspace}
\newcommand{\Li}{$^6$Li\xspace}

\usepackage{hyperref}%

\usepackage{comment}
\usepackage{graphicx}%
\usepackage{dcolumn}%
\usepackage{bm}%
\usepackage{xcolor} 

\usepackage{siunitx}
\DeclareSIUnit\bohr{\text {\ensuremath {a}}_{0}}
\DeclareSIUnit\gauss{\text{G}}
\usepackage{xspace}
\usepackage{braket}
\usepackage{placeins}

\usepackage{bm}
\usepackage{physics}
\usepackage{mathtools}
\usepackage{amsmath,amsfonts,amssymb}
\usepackage{bbold}

\usepackage{hyperref}%

\begin{document}

\title{Anderson orthogonality scaling in the Rabi-driven heavy Fermi polaron}  %

\author{Michael Rautenberg}
\thanks{These authors contributed equally to this work.}
\author{Tobias Krom}%
\thanks{These authors contributed equally to this work.}
\affiliation{%
 Physikalisches Institut, Universität Heidelberg, 69120 Heidelberg, Germany
}%

\author{Eugen Dizer}
\author{Olivier Bleu}
\affiliation{
 Institut für Theoretische Physik, Universität Heidelberg, 69120 Heidelberg, Germany
}%

\author{Eleonora Lippi}
\thanks{present address: Department of Physics and Research Center OPTIMAS, RPTU University Kaiserslautern-Landau, 67663 Kaiserslautern, Germany}
\affiliation{%
 Physikalisches Institut, Universität Heidelberg, 69120 Heidelberg, Germany
}%

\author{Tilman Enss}
\author{Manfred Salmhofer}
\affiliation{
 Institut für Theoretische Physik, Universität Heidelberg, 69120 Heidelberg, Germany
}%

\author{Lauriane Chomaz}
\email{chomaz@uni-heidelberg.de}
\author{Matthias Weidemüller}%
\email{weidemueller@uni-heidelberg.de}
\affiliation{%
 Physikalisches Institut, Universität Heidelberg, 69120 Heidelberg, Germany
}%

\date{\today}

\begin{abstract}

The Anderson orthogonality catastrophe (AOC) is a paradigmatic many-body phenomenon in which a local perturbation induces a macroscopic response of a Fermi sea. 
We probe signatures of the AOC by coherently driving heavy Fermi polarons in an ultracold \Li-\Cs mixture.  We observe a power-law dependence of the measured Rabi frequency on the drive strength, with exponents consistent with AOC predictions. %
Finite-temperature simulations quantitatively reproduce the observed scaling, indicating that AOC signatures persist beyond the idealized zero-temperature, infinite-mass limit. The damping of the Rabi oscillations provides access to polaron dephasing and reveals a nonmonotonic drive dependence, qualitatively consistent with current theories. 
Our results establish coherently driven impurities as a versatile probe of quantum many-body dynamics through local coherent control. %

\end{abstract}

\maketitle

A single impurity immersed in a non-interacting Fermi sea provides a paradigmatic setting for studying fermionic many-body systems~\cite{Bloch2008_Manybody}. The emergent Fermi polaron describes a mobile impurity in terms of a quasiparticle with modified properties~\cite{Landau1933_Uber, Landau1948_Effective, Parish2025_Fermi, Massignan2026_Polarons}. Anderson found an analytical solution for the limit of an infinitely heavy impurity~\cite{Anderson1967_Infrared}, where Landau Fermi-liquid theory breaks down. In this regime, the impurity becomes equivalent to a static scattering potential for the surrounding fermions. Remarkably, in the thermodynamic limit, an infinite number of elementary particle-hole excitations of the Fermi sea are needed to accommodate the impurity, thus rendering the many-body wavefunction orthogonal to the case without the impurity. This surprising result has been termed the ``Anderson Orthogonality Catastrophe" (AOC) and plays a key role in different contexts ranging from Fermi-edge singularities in X-ray spectra to transport processes in quantum dots~\cite{Mahan1967_Excitons,
Nozieres1969_Singularities, Ohtaka1990_Theory, Sankar2024_Detectortuned}. 

While ultracold atoms %
have enabled broad and detailed studies of Fermi polarons~\cite{Schirotzek2009_Observation,Kohstall2012_Metastability,Ness2020_Observation, Scazza2017_Repulsive}, signatures of the AOC have so far been elusive. Here, the Fermi sea %
consists of an ensemble of optically trapped, spin-polarized neutral fermionic atoms, while the impurities are realized either by a few atoms in a different internal state, or by a dilute sample of another atomic species, providing access to equal- and unequal-mass polarons%
~\cite{Massignan2014_Polarons, Parish2025_Fermi, Massignan2026_Polarons, Scazza2022_Repulsive, Baroni2024_Quantum}. %
Standard probes of the Fermi polaron include time- and frequency-domain spectroscopy. Heavy polaron spectra are well described by static-impurity theories, yet direct signatures of the AOC are masked by thermal effects~\cite{Cetina2016_Ultrafast, KromRautenberg2026_spectroscopy, RautenbergKrom2026_ramsey}. 
Alternatively, the polaron can be probed by its response to a coherent drive ~\cite{Kohstall2012_Metastability, Scazza2017_Repulsive, DarkwahOppong2019_Observation}.
Coherent driving over a broad range of drive strengths has recently been applied to equal-mass polarons~\cite{Vivanco2025_strongly}, whereas its application to heavy impurities is predicted to provide direct access to signatures of the AOC~\cite{Adlong2021_Signatures}.

\begin{figure}
    \centering
    \includegraphics[width=1\linewidth]{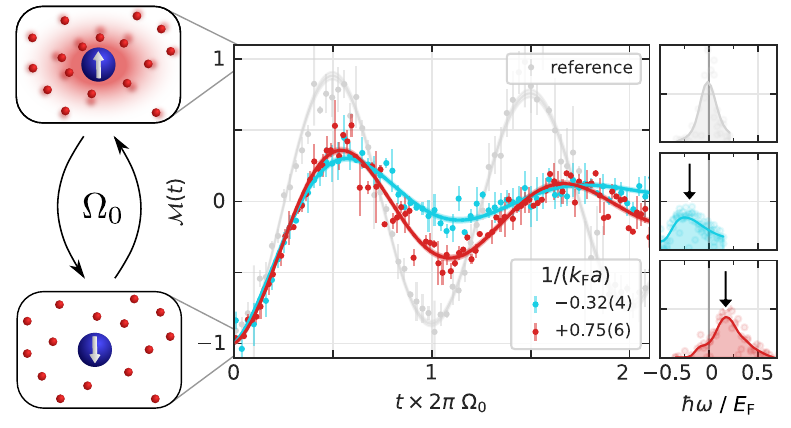}
    \caption{Left panel: Coherently driving a \Cs impurity (blue sphere) between a non-interacting reference state ($\ket{\downarrow}\equiv\mathrm{Cs}\ket{2}$, bottom left illustration, magnetization $\mathcal{M}=-1$) and an interacting state ($\ket{\uparrow}\equiv\mathrm{Cs}\ket{1}$, top left, $\mathcal{M}=+1$) that features tunable interactions with the surrounding fermionic \Li bath (red dots). The coupling strength is characterized by the Rabi frequency $\Omega_0$ of \Cs without \Li Fermi sea. Middle panel: The system response is extracted by fitting a damped oscillation (solid lines) with frequency $\Omega$ and damping rate $\Gamma$ to the oscillations of the impurity magnetization $\mathcal{M}(t)$ for different impurity-bath interactions parametrized by the parameter $1/(k_\mathrm{F} a)$ with the Fermi wavenumber $k_\mathrm{F}$ and the impurity-bath scattering length $a$ (gray: reference oscillation without Fermi sea providing $\Omega_0 = 2\pi\times \SI{7.62(2)}{\kilo\hertz}\approx\num{0.2}E_\mathrm{F}/\hbar$, where $E_\mathrm{F}$ is the Fermi energy).  Right panel: Shown are the corresponding injection spectra taken in the linear-response regime. The arrows indicate the energy of the drive, which is chosen to be resonant with the attractive (cyan) or repulsive (red) polaron branch, respectively. 
    }
    \label{fig:Fig1_Rabi}
\end{figure}

In this Letter, we investigate the response of a heavy Fermi polaron in an ultracold Fermi gas to a continuous, resonant external drive for an impurity with the largest mass imbalance accessible in alkali mixtures, as shown in Fig.~\ref{fig:Fig1_Rabi}. %
We find that the oscillation frequency as a function of the drive strength follows a power-law scaling, with the corresponding exponent closely matching the Fermi-edge singularity exponent predicted by the AOC for different interaction strengths. Numerical simulations of a coherently driven static impurity at finite temperature agree well with the measured Rabi frequencies, but describe the magnitude of the measured damping rates less accurately.

The qualitative difference between coherently driven \textit{mobile} and \textit{static} impurities can be pictured as follows. For a \textit{mobile} impurity, resonantly driving between the polaron state and a non-interacting reference state %
provides access to the quasiparticle residue $Z$~\cite{Kohstall2012_Metastability, Scazza2017_Repulsive, DarkwahOppong2019_Observation}, which quantifies the wavefunction overlap of these two states~\cite{Massignan2014_Polarons}. %
Within the variational ansatz introduced by Chevy~\cite{Chevy2006_Universal}, and for sufficiently weak drive strengths, %
the interacting Rabi frequency $\Omega$ 
relates to the bare impurity Rabi frequency $\Omega_0$, which measures the drive strength, via $\Omega/\Omega_0=\sqrt{Z}$.
Extensions to this ansatz, explicitly including the external drive into the impurity dynamics, predict a more intricate dependence of $\Omega/\Omega_0$ on the drive strength~\cite{Adlong2020_Quasiparticle, Hu2023_Fermi, Mulkerin2024_Rabi}. In particular, a finite polaron spectral width $\Gamma_0$ in the absence of the external drive modifies the weak-drive result to $\Omega/\Omega_0 = \sqrt{Z - \Gamma_0^2/\Omega_0^2}$~\cite{Adlong2020_Quasiparticle}. %
For sufficiently small $\Gamma_0$, there is a range of drive strengths where $\Omega/\Omega_0$ plateaus, allowing to extract $Z$. 

Conversely, for an infinitely heavy, i.e., \emph{static}, impurity, a  characteristic power-law reduction of the Rabi frequency %
\begin{equation} \label{eq:AOC_scaling}
    \frac{\Omega}{\Omega_0} \propto \left(\frac{\hbar\Omega_0}{E_\mathrm{F}}\right)^{\alpha/(2-\alpha)}  %
\end{equation}
has been derived for zero temperature~\cite{Knap2013_Dissipative}, as well as numerically studied for finite temperatures and drive strengths $\hbar \Omega_0$ up to the Fermi energy $E_\mathrm{F}$~\cite{Adlong2021_Signatures}.
In contrast to the mobile impurity case, $\Omega/\Omega_0$ exhibits no drive-independent plateau, reflecting the vanishing quasiparticle residue expected from the AOC. This scaling is universal in the sense that its exponent depends only on the scattering phase shift at the Fermi surface $\delta(k_\mathrm{F})$, which in our case depends only on the impurity-bath interactions, see Supplementary Material (SM)~\cite{suppl}, via $\alpha=\delta(k_\mathrm{F})^2/\pi^2$ \footnote{This is the relevant exponent for the cases studied here; for repulsive interactions and a drive set to the \emph{attractive} polaron branch, one would expect $\tilde{\alpha} = \left[\delta(k_\mathrm{F})/\pi + 1 \right]^2$ instead~\cite{Knap2013_Dissipative, Adlong2021_Signatures}.}, and is a direct consequence of the Fermi-edge singularity. Here we probe this scaling in experiment.

To study the heavy Fermi polaron,  
we prepare a sample of $\lesssim\num{1e3}$ \Cs impurities in a degenerate Fermi gas of $\approx\num{1.8e5}$ spin-polarized \Li atoms, where the impurities are centered in an almost constant density regime of the surrounding Fermi gas with a local Fermi energy $E_\mathrm{F} = h \times \SI{31(3)}{\kilo\hertz} = k_\mathrm{B} \times \SI{1.49(15)}{\micro\kelvin}$~\cite{KromRautenberg2026_exp, suppl}. The impurities are thermalized to the fermionic bath at a reduced temperature of $T/T_\mathrm{F} = \num{0.25(3)}$ and are prepared in the effectively non-interacting reference state Cs$\ket{2}$~\cite{suppl}. Optical Raman spectroscopy is used to drive the impurities to the Cs$\ket{1}$ state, as schematically shown on the left of Fig.~\ref{fig:Fig1_Rabi}, featuring tunable interactions with the surrounding bath provided by a magnetic Feshbach resonance close to $\SI{888.6}{\gauss}$~\cite{Repp2013_Observation, Pires2014_Analyzing, KromRautenberg2026_exp, suppl}. We parametrize impurity-bath interactions with the interaction parameter $1/(k_\mathrm{F}a)$, where $a$ is the Cs$\ket{1}$-Li $s$-wave scattering length and $1/k_\mathrm{F}=\SI{3110+-150}{\bohr}$ is the inverse Fermi momentum \cite{suppl}.

The Raman spectroscopy is set up to not transfer any momentum, equivalent to radio-frequency spectroscopy, and allows us to reach bare Cs Rabi frequencies up to $\Omega_0/(2\pi) \approx \SI{500}{\kilo\hertz} \gg E_\mathrm{F}/h$, controlled by the intensities of the Raman lasers, thus enabling the study of the driven polaron across all relevant regimes. %
The energy of the coherent drive is set by the two-photon Raman detuning $\hbar \omega$ from the bare Cs transition at vanishing impurity-bath interactions (reference spectrum in gray in Fig.~\ref{fig:Fig1_Rabi}). 
Throughout this work, we set the energy of the drive to match the polaron peak energy in the absence of Rabi drive $\hbar \omega = E_\mathrm{p}$, which is determined using linear-response injection spectroscopy (marked by the arrows in the spectra in Fig.~\ref{fig:Fig1_Rabi}). 
On the attractive side of the resonance, we set the drive energy to the attractive polaron peak, whereas on the repulsive side we tune it to the repulsive polaron energy.

\begin{figure}
    \centering
    \includegraphics[width=1\linewidth]{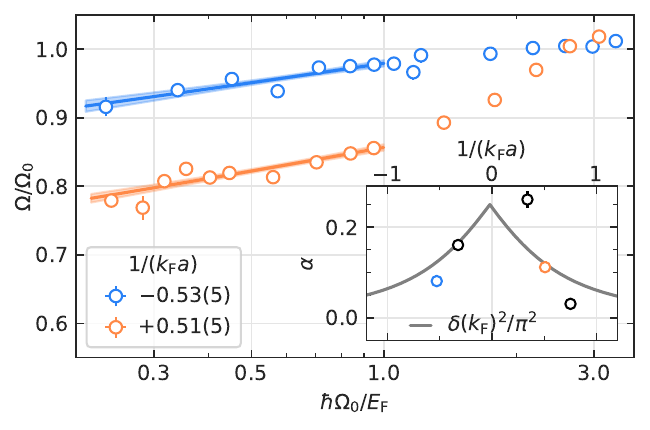}
    \caption{Response of the driven heavy Fermi polaron as a function of drive strength $\hbar\Omega_0$ for two exemplary impurity-bath interactions $1/(k_ \mathrm{F}a)$. %
    On the attractive side of the Feshbach resonance ($1/(k_ \mathrm{F}a)<0$)
    the response of the \emph{attractive} polaron is shown, whereas we probe the \emph{repulsive} polaron branch on the repulsive side.
    Error bars indicate fit uncertainties.
    The solid lines are fits of Eq.~(\ref{eq:AOC_scaling}) used to extract the scaling exponent $\alpha$, where the shaded areas show the corresponding $1\sigma$ confidence band of the fits. The fit results are shown in the inset with matching colors (data for the remaining interactions can be found in the SM~\cite{suppl}) together with the universal behavior $\alpha=\delta(k_\mathrm{F})^2/\pi^2$. Vertical error bars again indicate fit uncertainties.%
    }
    \label{fig:Rabi_AOC_fit}
\end{figure}

As shown in Fig.~\ref{fig:Fig1_Rabi}, we measure the temporal evolution of $N_1$ and $N_2$, the population in states Cs$\ket{1}$ and Cs$\ket{2}$, respectively, using absorption imaging and compute the impurity magnetization $\mathcal{M} = (N_1-N_2)/(N_1+N_2)$ (colored data points in Fig.~\ref{fig:Fig1_Rabi}). 
The response of the driven polaron is then obtained by fitting the damped oscillations of the impurity magnetization $\mathcal{M}(t)$ as a function of the drive time $t$ to extract the interacting Rabi frequency $\Omega$ and damping rate $\Gamma$~\cite{suppl}. %
The bare Cs Rabi frequency $\Omega_0$ is measured analogously from a reference measurement without the surrounding Li bath (shown in gray). At the weak drive strengths used here, also the bare Cs Rabi oscillations are damped, mainly because of magnetic field fluctuations. 
More details on the systematics of the Raman driving can be found in Ref.~\cite{KromRautenberg2026_exp}.

We note that the total number of Cs atoms remains approximately constant throughout the oscillations, while the damping rate in the absence of the Li bath is about one order of magnitude smaller than in its presence.
Thus, we conclude that the observed damping of the interacting Rabi oscillations is dominated neither by finite lifetimes nor by experimental decoherence \cite{suppl}. Instead, it reflects genuine dephasing of the driven many-body state. The measurements throughout this work are taken over a range of drive strengths from $\hbar\Omega_0 \approx 0.2E_\mathrm{F}$ to $\hbar\Omega_0 > 3E_\mathrm{F}$, as for lower drive strengths the oscillations become overdamped, see below.

In Fig.~\ref{fig:Rabi_AOC_fit}, we investigate the reduction of the Rabi frequency $\Omega$ of the interacting polaron compared to the bare Cs Rabi frequency $\Omega_0$ for two exemplary impurity bath interactions \footnote{A differential AC Stark shift induced by the Raman coupling causes the Feshbach resonance position to vary linearly with $\hbar \Omega_0$.
This results also in an approximately linear shift of the interaction parameter $1/(k_\mathrm{F}a)$ by \num[retain-explicit-plus=true]{+0.04(1)} per $E_\mathrm{F}$ towards stronger repulsion~\cite{suppl}. %
Since all relevant polaron properties vary only weakly as a function of interactions, this effect is negligible for drive strengths $\hbar \Omega_0\lesssim E_\mathrm{F}$.}. %
We observe a mostly monotonic increase of $\Omega/\Omega_0$ for increasing drive strengths for both attractive and repulsive polaron branches. In the limit of strong drive $\hbar\Omega_0\gg E_\mathrm{F}$, the system recovers the expected two-level dynamics $\Omega \to \Omega_0$. We find small overshoots where $\Omega > \Omega_0$, as also reported in~\cite{Vivanco2025_strongly} and further discussed in the SM~\cite{suppl}, where we also show the response of the driven polaron for three additional interaction parameters.

We do not find a pronounced drive-independent plateau of $\Omega/\Omega_0$ at weak drive strengths, in contrast to earlier findings in equal-mass~\cite{Vivanco2025_strongly} and other mass-imbalanced polaron systems~\cite{Kohstall2012_Metastability}. 
Instead, as shown by the solid lines in Fig.~\ref{fig:Rabi_AOC_fit}, the behavior of $\Omega/\Omega_0$ is consistent with a power-law scaling for $k_\mathrm{B}T \lesssim \hbar\Omega_0 < E_\mathrm{F}$, one of the predicted signatures of the AOC~\cite{Adlong2021_Signatures}. %
In this range of drive strengths, we fit the exponent of the AOC scaling law in Eq.~(\ref{eq:AOC_scaling}) and compare the results to the expected exponent given by the scattering phase shift at the Fermi surface for different interaction parameters, see inset of Fig.~\ref{fig:Rabi_AOC_fit}. Overall, we find qualitative agreement between the extracted power-law exponents and the theoretical predictions. We note that small quantitative deviations are expected due to finite temperature effects~\cite{Adlong2021_Signatures, suppl}.

These findings indicate a surprising robustness of the AOC scaling law against finite temperature and mass imbalance, both of which in principle modify the AOC scaling~\cite{Schmidt2018_Universal,Chen2025_MassGap,Ramos2026_MassGapFDA}. 
To understand the effect of finite mass and temperature, we compare the measured response of the driven Fermi polaron to three complementary theoretical approaches, for details see the SM~\cite{suppl}. The first is a ``coupled FDA'' model based on the functional determinant approach (FDA)~\cite{Knap2012_TimeDependent, Schmidt2018_Universal, Wang2023_Functional, Drescher2024_Bosonic}. It treats the impurity as infinitely heavy and effectively includes infinitely many particle-hole excitations, thereby capturing the full AOC physics. The second is a ``coupled T-matrix'' model based on the non-self-consistent $T$-matrix approach~\cite{Chevy2006_Universal, Parish2016_Quantum}. It retains the finite impurity-to-bath mass ratio but restricts the dressing to a single particle-hole excitation of the Fermi sea. Both models incorporate the drive solely through a $2\times2$ Rabi-coupled impurity Green's function that connects the interacting and non-interacting states~\cite{suppl}. We additionally compare to a more sophisticated ``full $T$-matrix'' approach~\cite{Hu2023_Fermi,Mulkerin2024_Rabi,Vivanco2025_strongly}, in which the drive also enters the self-energy.
The coupled FDA and coupled $T$-matrix are good approximations in the weak-drive regime, whereas the full $T$-matrix also captures the strong-drive limit. All models are evaluated at the experimental temperature of $T/T_\mathrm{F}=0.25$.

\begin{figure}
    \centering
    \includegraphics[width=.9\linewidth]{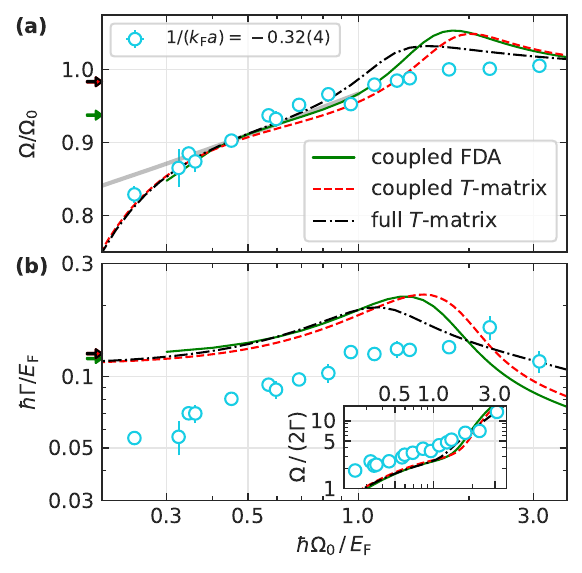}
    \caption{Theory comparison of the response of the driven attractive Fermi polaron as a function of drive strength at attractive impurity-bath interactions $1/(k_\mathrm{F}a) = \num{-0.32(4)}$.
    We show the experimentally measured reduction of Rabi frequencies (\textbf{a}) and damping rates (\textbf{b}) together with their predictions from different numerical simulations, see main text and~\cite{suppl}. The scaling prediction of Eq.~(\ref{eq:AOC_scaling}) is shown as solid gray line where the prefactor has been adjusted to match the data.
    The arrows on the left show predictions for the quasiparticle weight $\sqrt{Z}$ (a) and spectral half width at half maximum $\Gamma_0$ (b) in the absence of external drive, calculated within $T$-matrix (red and black) and FDA (green) theories at the experimental temperature.
    The inset shows the ratio $ \Omega/(2\Gamma)$.
    }
    \label{fig:Rabi_theory_comparison}
\end{figure}

In Fig.~\ref{fig:Rabi_theory_comparison}, we compare the theory predictions to one set of measurements on the attractive side of the Feshbach resonance 
\footnote{%
We also observe coherent oscillations on the repulsive polaron branch that exhibit a scaling behavior compatible with Eq.~(\ref{eq:AOC_scaling}), \textit{cf.} Fig.~\ref{fig:Rabi_AOC_fit}. In this parameter regime all numerical models predict more intricate dynamics of $\mathcal{M}(t)$, rather than a damped oscillation characterized by a single frequency, due to the presence of the additional polaron branch and bound state \cite{suppl}.}. 
The detuning of the Rabi drive is set to match the attractive polaron energy. For the reduction of Rabi frequencies (Fig.~\ref{fig:Rabi_theory_comparison}a), we find good quantitative agreement with the coupled FDA calculations for drive strengths $\hbar\Omega_0 \lesssim E_\mathrm{F}$ \footnote{Due to approximations in the model, the results of the coupled FDA and coupled $T$-matrix calculations become increasingly inaccurate at large $\hbar\Omega_0 \gtrsim E_\mathrm{F}$, \textit{cf.}~\cite{suppl}} (solid green line). The FDA results in turn agree with the scaling law Eq.~\eqref{eq:AOC_scaling}, shown as the solid gray line. We test this agreement numerically for various temperatures $0\leq T\leq 0.25\,T_\mathrm{F}$ in the SM~\cite{suppl}. With increasing temperature, the range of drive strengths over which the scaling law is visible shrinks, but the scaling exponent changes only weakly \cite{Adlong2021_Signatures}. %

Surprisingly, the $T$-matrix calculations (red dashed and black dash-dotted lines) show a very similar behavior to FDA, indicating that higher-order particle-hole excitations do not significantly affect the Rabi dynamics in this regime. Indeed, even within $T$-matrix theories, including only a single particle-hole excitation, we numerically observe power-law behavior over a limited range of drive strengths. This behavior emerges as the impurity-bath mass ratio is increased to $M/m\gtrsim 10$ at our experimental temperatures~\cite{suppl}. 
At lower temperatures, where the AOC scaling law extends to weaker drive strengths, deviations between $T$-matrix and FDA calculations arise~\cite{suppl}.

On the other hand, the damping rate of the Rabi oscillations is overestimated by all the theoretical models for weak drive strengths (Fig.~\ref{fig:Rabi_theory_comparison}b).  The different theories agree closely with each other for $\hbar\Omega_0\lesssim E_\mathrm{F}$ and converge to the quasiparticle width $\Gamma_0$ in the absence of external drive (arrows on the left of Fig.~\ref{fig:Rabi_theory_comparison}b). The comparably strong damping predicted by the different theories prevents $\Omega/\Omega_0$ to converge to the quasiparticle weight $\sqrt{Z}$ (arrows in Fig.~\ref{fig:Rabi_theory_comparison}a), and restricts the drive-strength range over which the scaling law is observable, see SM~\cite{suppl}. %
Experimentally, the low damping rates imply that the modified Rabi frequencies are not dominated by damping over the reported drive-strength range, see inset of Fig.~\ref{fig:Rabi_theory_comparison}b and SM~\cite{suppl}. %
The oscillations become overdamped only for $\hbar\Omega_0\lesssim0.2E_\mathrm{F}$, %
while the quality of the oscillations improves with increasing drive strength. %

\begin{figure}
    \centering
    \includegraphics[width=1\linewidth]{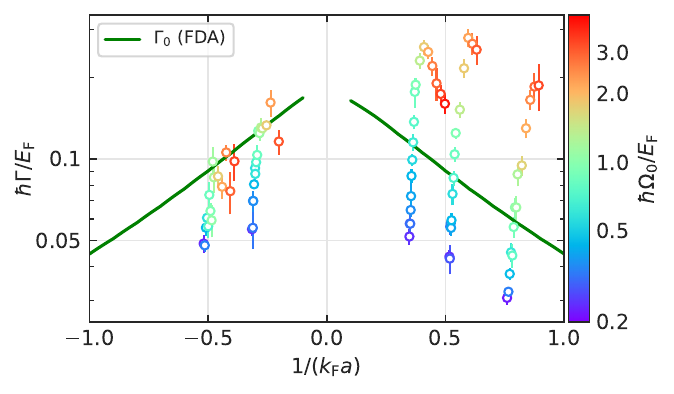}  %
    \caption{Dephasing rates measured as the damping of Rabi oscillations as a function of drive strength (color coded) for different interactions for the attractive ($a<0$) and repulsive ($a>0$) polaron, respectively. The solid line is the FDA prediction for the spectral width (half width at half maximum) in the absence of external drive. 
    A more detailed presentation of the data across different interactions and drive strengths can be found in the SM~\cite{suppl}.
    }
    \label{fig:Rabi_damping}
\end{figure}

Besides influencing the reduction of Rabi frequencies, the damping rates of the oscillations also reveal interesting physical properties of the driven polaron ~\cite{Scazza2022_Repulsive, Massignan2011_Repulsive} as they can be interpreted as a dephasing inherent to the driven many-body system~\cite{suppl}. 
At any fixed interaction strength, the damping first increases with increasing drive (Fig.~\ref{fig:Rabi_theory_comparison}b and \cite{suppl}) before it decreases again for very strong drives $\hbar\Omega_0 \gg E_\mathrm{F}$, \textit{cf.}~\cite{Shkedrov2022_Absence, Vivanco2025_strongly}. The dephasing rate of the Rabi oscillations at low drive corresponds to neither the theoretically predicted nor the experimentally measured \cite{KromRautenberg2026_spectroscopy} spectral width of the polaron spectrum at the corresponding interactions (\textit{cf.}~green arrow in Fig.~\ref{fig:Rabi_theory_comparison}b). This is in contrast to theoretical predictions~\cite{Adlong2020_Quasiparticle} and earlier observations in equal mass systems~\cite{Scazza2017_Repulsive, DarkwahOppong2019_Observation} using drive strengths $\hbar\Omega_0/E_\mathrm{F} \approx 0.7 \dots 1.7$~\footnote{Refs.~\cite{Scazza2017_Repulsive, DarkwahOppong2019_Observation} measured the damping rates at drive strengths $\hbar\Omega_0/E_\mathrm{F} = 0.7$ and $0.9\dots1.7$, respectively. For these drive strengths, we also find reasonable agreement with the FDA spectral width.}. Using lower drive strengths, dephasing rates of roughly half of the measured spectral width have also been reported in an equal-mass system~\cite{Vivanco2025_strongly}.

The mismatch between dephasing rates at small drive strengths and spectral widths persists across all interactions explored in this work, see Fig.~\ref{fig:Rabi_damping}. Furthermore, 
an asymmetry between attractive and repulsive interactions can be observed for the strongly driven case $\hbar\Omega_0 \gtrsim E_\mathrm{F}$, where the maximal decoherence rates for the repulsive polaron branch exceed those of the attractive one. %
Differences in damping between the two polaron branches have been attributed to the difference between many-body dephasing and momentum relaxation~\cite{Mulkerin2024_Rabi, Wasak2024_Decoherence}. In our case, however, momentum relaxation should be strongly suppressed due to the large impurity-bath mass ratio. 
An additional source of the observed branch asymmetry may be that, unlike the attractive polaron, the repulsive polaron is intrinsically metastable, being embedded in the molecule–hole continuum that lies above the attractive-polaron ground state even in the absence of driving~\cite{Massignan2014_Polarons, Adlong2020_Quasiparticle}. Strong driving pushes one of the dressed branches toward this continuum, enhancing hybridization and damping~\cite{Vivanco2025_strongly}.

In summary, we find that the measured Rabi frequencies at low drive strengths compare favorably with the power-law scaling predicted by the AOC. 
Our observations suggest that signatures of the AOC are substantially more accessible in dynamical impurity probes than expected from the strict thermodynamic and static-impurity limits in which it was originally formulated. 
Indeed, the observed scaling emerges despite the finite impurity-to-bath mass ratio $M/m\approx22$, nonzero temperature $T/T_\mathrm{F}\approx 0.25$, and finite system size, all of which yield a finite quasiparticle residue~\cite{Schmidt2018_Universal,Chen2025_MassGap,Ramos2026_MassGapFDA} and should in principle modify the orthogonality-catastrophe scaling. 
Together with extensive numerical studies, tracking the build-up of the AOC power law in the Rabi frequency of the driven Fermi polaron as a function of mass ratio and temperature, our measurements establish the observation of a finite-temperature, finite-drive precursor of AOC scaling---one that already closely approaches the predicted asymptotic power law.

Experimentally, continuously driving the polaron offers an advantage because it probes the impurity self-energy at a new, externally imposed energy scale set by the drive strength, thereby circumventing the thermal decoherence that obscures AOC signatures in the low-energy regime. Furthermore, for Rabi driving, the scaling exponent enters through the frequency, while for Ramsey interferometry \cite{RautenbergKrom2026_ramsey} and linear-response spectroscopy \cite{KromRautenberg2026_spectroscopy},  it affects only the amplitude of the measured signals, which is more challenging to measure precisely.

Finally, we observe a nonmonotonic dependence of the dephasing rates of the drive-dressed heavy Fermi polaron on drive strength. Especially the weak-drive regime, captured by neither static- nor mobile-impurity theories, provides a benchmark for future many-body theories.

More generally, our work highlights how coherent control of localized impurities can provide direct access to the collective many-body response of an interacting fermionic medium.
The high degree of tunability available in ultracold mixtures offers promising avenues for exploring extensions ranging from dynamically screened impurity interactions and spin-exchange couplings to non-equilibrium orthogonality catastrophe and quantum-transport phenomena~\cite{Knap2012_TimeDependent, You2019_Atomtronics}. 

We thank F.~Scazza, N.~Navon, R.~Schmidt, R.~Grimm, C.~Baroni, E.~Dobler, G.~Roati, M.~Zaccanti, S.~Jochim, and G.~Zürn for fruitful discussions.
This work is funded by the Deutsche Forschungsgemeinschaft DFG (German Research Foundation) under Project-ID 273811115 – SFB 1225 ISOQUANT, and the Heidelberg Excellence Cluster STRUCTURES (EXC 2181/1 - 390900948). Support by the Heidelberg Center for Quantum Dynamics is gratefully acknowledged. M.R., T.K, and E.L. acknowledge support by the International Max Planck Research School for Quantum Dynamics in Physics, Chemistry and Biology (IMPRS-QD). M.R. acknowledges financial support from the German Academic Exchange Service (DAAD).

\FloatBarrier
\bibliography{main}

\end{document}

\title{Supplemental Material for \\``Anderson orthogonality scaling in the Rabi-driven heavy Fermi polaron"}

\author{Michael Rautenberg}
\thanks{These authors contributed equally to this work.}
\author{Tobias Krom}%
\thanks{These authors contributed equally to this work.}
\affiliation{%
 Physikalisches Institut, Universität Heidelberg, 69120 Heidelberg, Germany
}%

\author{Eugen Dizer}
\author{Olivier Bleu}
\affiliation{
 Institut für Theoretische Physik, Universität Heidelberg, 69120 Heidelberg, Germany
}%

\author{Eleonora Lippi}
\thanks{present address: Department of Physics and Research Center OPTIMAS, RPTU University Kaiserslautern-Landau, 67663 Kaiserslautern, Germany}
\affiliation{%
 Physikalisches Institut, Universität Heidelberg, 69120 Heidelberg, Germany
}%

\author{Tilman Enss}
\author{Manfred Salmhofer}
\affiliation{
 Institut für Theoretische Physik, Universität Heidelberg, 69120 Heidelberg, Germany
}%

\author{Lauriane Chomaz}
\author{Matthias Weidemüller}%
\affiliation{%
 Physikalisches Institut, Universität Heidelberg, 69120 Heidelberg, Germany
}%

\date{\today}

\maketitle

\tableofcontents

\newpage

\section{Experimental methods}
In this Section we describe the experimental details concerning the characterization of the sample as well as the procedures used to extract the Rabi frequencies and damping rates discussed in the main text. General details about sample preparation and experimental techniques can be found in Ref.~\cite{KromRautenberg2026_exp}. 

All measurements start from a sample of  $\lesssim\num{1000}$ \Cs impurities in a degenerate Fermi gas of $\approx\num{1.8e5}$ Li atoms, with all atoms of each species prepared in the same internal state.
Due to the much smaller spatial extent of the Cs cloud compared to the Li bath, the impurities sample a region of almost constant Li density (calculated as the mean Li density weighted by the normalized Cs density) $\overline{n}_\mathrm{Li} = \SI{3.8(6)e12}{\per\cubic\centi\meter}$. Similarly, we calculate the mean local Fermi energy $E_\mathrm{F} = h \times \SI{31(3)}{\kilo\hertz} = k_\mathrm{B} \times \SI{1.49(15)}{\micro\kelvin}$, resulting in the values reported in the main text. The impurities are thermalized to the fermionic bath at a reduced temperature of $T/T_\mathrm{F} = \num{0.25(3)}$. The small maximum relative impurity density $n_\mathrm{imp} = n_\mathrm{0,\,Cs} \, / \, \overline{n}_\mathrm{Li} = \num{0.15(4)}$, where $n_\mathrm{0,\,Cs}=\SI{6(1)e11}{\per\cubic\centi\meter}$ is the peak Cs density, justifies the interpretation of impurity physics (these numbers are calculated for $1000$ Cs impurities, in most data sets the actual Cs number is closer to half of that). Uncertainties include slow drifts of the Li atom number and temperature on a time scale of $\sim24$ hours.

We work at magnetic offset fields close to the Li$\ket{2}$-Cs$\ket{1}$ Feshbach resonance near \SI{888.6}{\gauss}, that allows the tuning of the Li$\ket{2}$-Cs$\ket{1}$ $s$-wave scattering length $a$. Internal states of Li and Cs are labeled according to their energy at finite magnetic field, starting with the ground state $\ket{1}$, followed by $\ket{2}$, etc. 
Using the inverse Fermi wavevector $1/k_\mathrm{F}=\SI{3110+-150}{\bohr}$ ($a_0$ is the Bohr radius), this scattering length defines the interaction parameter $1/(k_\mathrm{F}a)$ used in the main text to characterize impurity-bath interactions  . At the same magnetic fields, the Cs$\ket{2}$ state is only weakly interacting with the surrounding bath, with an approximately constant scattering length $a_\mathrm{Li\ket{2}-Cs\ket{2}} \approx \SI{-41}{\bohr}$.

The AOC scaling exponent (\textit{cf.} Eq.~(1) of the main text), is given by $\alpha=\delta(k_\mathrm{F})^2/\pi^2$.
The scattering phase shift $\delta(k)$ at a wavevector $k$ is defined via $-\cot \delta(k) = 1/(k a) + R^* k$. 
In the present case of the broad Li$\ket{2}$-Cs$\ket{1}$ Feshbach resonance near \SI{888.6}{\gauss}, the second part of this expression is negligibly small, because of the small range parameter $R^*=\SI{65}{\bohr}\ll1/k_\mathrm{F}$~\cite{Johansen2017_Testing, Ulmanis2015_Universality}.

\begin{figure}[t!]
    \centering
    \includegraphics[width=0.45\linewidth]{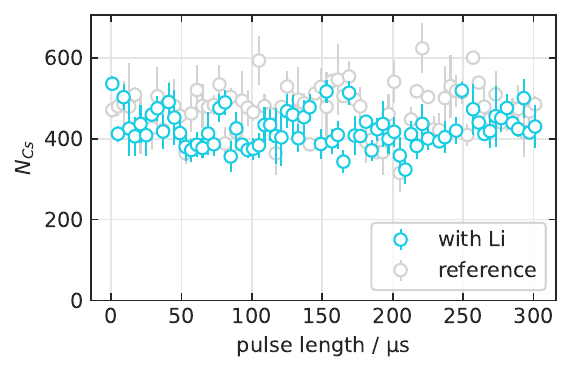}
    \caption{Example of the total cesium atom number $N_{Cs} = N_1 + N_2$ as a function of time during which the external drive is applied at $1/(k_\mathrm{F}a) = -0.32(4)$, corresponding to the cyan data set in Fig.~1 of the main text. The gray data points show the same measurement in the absence of the Li Fermi sea. Markers indicate the mean atom numbers over three independent measurements with error bars indicating the standard error of the mean. In both cases, no systematic reduction can be observed during the measurements. }
    \label{fig:N_Cs}
\end{figure}

\paragraph*{Coherence.}
In the main text we interpret the damping of Rabi oscillations as pure many-body dephasing, \textit{i.e.} we assume negligible atom loss and technical dephasing during the Rabi measurements. The following considerations support these statements. 
Inelastic two-body collision timescales involving excited Cs atoms or any three-body collision timescales are estimated to be slower than tens of milliseconds~\cite{KromRautenberg2026_exp}. Therefore, we do not observe any atom loss during the Rabi oscillation measurements, as exemplified in Fig.~\ref{fig:N_Cs}.
Furthermore, an inhomogeneous driving field, magnetic offset field, or Li density could in principle result in additional decoherence.
The Rabi coupling between Cs$\ket{1}$ and Cs$\ket{2}$ is realized via a two-photon Raman transition without momentum transfer (bare energy difference between the two states is $\approx h\times\SI{260}{\mega\hertz}\gg E_\mathrm{F}, k_\mathrm{B}T$). The Raman beams are much larger (beam waist $w \approx \SI{750}{\micro \meter}$) than the atomic clouds (all Cs $1/e$-radii $\sigma_i\lesssim\SI{10}{\micro\meter}$) and carefully aligned, resulting in a homogeneous driving field. Also the magnetic field does not vary on the scale of a few tens of micrometers explored by the impurities~\cite{KromRautenberg2026_exp}. Lastly, as stated above, the Li density is homogeneous, so that $E_\mathrm{F}$ varies by less than \SI{15}{\percent} across the Cs cloud.
We attribute the fact that we still observe damping on the bare Cs Rabi oscillations (without surrounding Fermi sea) mainly to shot-to-shot fluctuations of the magnetic field ($\sim \SI{10}{\milli\gauss}$) and to a lesser extent to laser intensity fluctuations (on the level of a few percent)~\cite{KromRautenberg2026_exp}.
The observed bare Cs damping rates in the absence of the Li bath are about one order of magnitude smaller than in its presence.

\subsection{Analyzing measured Rabi oscillations}
\label{sec:fit_routines}

To extract the response of the driven heavy Fermi polaron, we fit the time evolution of the magnetization $\mathcal{M} = (N_1-N_2)/(N_1+N_2)$ with a model of the form 
\begin{equation} \label{eq:fit_model}
    \mathcal{M}(t) = \widetilde{\mathcal{M}}-A \cos(\Omega t) e^{-\Gamma t} - (1 - A +\widetilde{\mathcal{M}}) e^{-\Gamma t} \ ,
\end{equation}
where $A$ is a dimensionless amplitude and $\Gamma$ a damping rate. For long times, the magnetization approaches its steady-state value $\widetilde{\mathcal{M}}$ where the last term ensures that $\mathcal{M}(t=0) = -1$ regardless of the fit parameters, reflecting the initial conditions that we always prepare the sample in the non-interacting impurity state Cs$\ket{2}$. To calibrate the bare Cs Rabi frequency $\Omega_0$, we perform the same fit on the response of the Cs atoms without the surrounding Li bath, as exemplified in the central panel of Fig.~1 in the main text. For the measurements presented in the main text, we record these reference measurements without Li for every data point (i.e. impurity-bath interaction and drive strength) to get a reliable measurement of $\Omega/\Omega_0$. %

A more general fitting model with two different decay rates replacing $\Gamma$ in the second and third term of Eq.~(\ref{eq:fit_model}) has been used in the literature \cite{Scazza2017_Repulsive, DarkwahOppong2019_Observation, Vivanco2025_strongly}. Since in the measurements presented in this work the steady-state magnetization $\widetilde{\mathcal{M}}$ is always close to zero, this more general form does not change the extracted Rabi frequencies, which we explicitly verified on the data.

\subsection{Raman light-induced shift of the Feshbach resonance} \label{sec:FR_shift}

Since we use an optical Ramsey method to realize the coupling between the impurity states which relies on near resonant light, unwanted AC Stark shifts of the atomic states generating the Feshbach resonance are a concern. Indeed, we characterized this shift in \cite{KromRautenberg2026_exp}, where we find it to cause a change in interaction parameter of 
\begin{equation}
    (k_\mathrm{F}a)^{-1} = (k_\mathrm{F}a)^{-1}_0 + 0.04(1) \frac{\hbar \Omega_0}{E_\mathrm{F}}
\end{equation}
from the interaction parameter $(k_\mathrm{F}a)^{-1}_0$ without additional drive.

For all drive strengths up to $\approx E_\mathrm{F}$ this effect is expected to be negligible as all relevant quantities (detuning to the polaron energy, quasiparticle-residue, spectral widths, \textit{etc.}) vary slowly as a function of $(k_\mathrm{F}a)^{-1}$. 
Note that all statements about the AOC scaling law only use data for drive strengths $\hbar \Omega_0 \lesssim E_\mathrm{F}$. For strong drives $\hbar \Omega_0 \gg E_\mathrm{F}$ the interaction parameter shift indeed complicates the interpretation of our results, which however does not alter the conclusions drawn in the main text.

\section{Theoretical description}
\label{sec:model}

In this Section, we present the theoretical approaches used to calculate the renormalized Rabi frequencies and oscillation decay rates compared with the experimental data in the main text. We start by introducing the theories in the absence of Rabi drive in Sec. \ref{subsec:undriven} before discussing the problem in presence of the drive in Sec. \ref{subsec:driven_winter}.

We consider a single impurity in a Fermi gas with an interacting state $\ket{\up}$ and another internal state $\ket{\dn}$, which has negligible interactions with the bath. The Hamiltonian for the interacting impurity in a Fermi gas is given by
\begin{align}  \label{eq:H_mobile}
    \hat{H}=\sum_\kk (\epsilon_\kk-\mu) \cop^\dagger_\kk\cop_\kk
  +\sum_{\kk} \eI_\kk \dop^\dagger_{\kk\up}\dop_{\kk\up}
  +\frac{g}{\mathcal{V}} \sum_{\kk',\kk,\q} \dop^\dagger_{\kk'+\q \up}\dop_{\kk' \up}\cop^\dagger_{\kk-\q}\cop_\kk \,,
\end{align}
where $\cop_\kk$ ($\cop_\kk^\dagger$) are annihilation (creation) operators for fermions in the bath, and  $\dop_{\kk\up}$ ($\dop_{\kk\up}^{\dagger}$) are annihilation (creation) operators for the impurity in state $\ket{\up}$. $\mu$ is the chemical potential of the bath, and $\epsilon_\kk=\kk^2/(2m)$ and $ \epsilon^I_\kk=\kk^2/(2M)$ denote the kinetic energies of bath fermions and the impurity, respectively. Finally, $g$ corresponds to the ``bare'' strength of the contact interactions between medium particles and the impurity in state $\ket{\up}$. It is related to the corresponding $s$-wave scattering lengths $a$ via $\frac{1}{g} = \frac{m_r}{2\pi a} - \sum_{\kk}^{\Lambda} \frac{1}{\epsilon_\kk+\epsilon^I_\kk}$ with reduced mass $m_r=M m/(M+m)$, and $\Lambda$ an ultraviolet cutoff which will eventually be sent to infinity at the end of the calculation. 

\newpage

We consider the scenario where the impurity is coherently driven between the interacting state $\ket{\up}$ and the non-interacting state $\ket{\dn}$ by a field of bare Rabi frequency $\Omega_0$ and detuning $\Delta_0$. This is modeled by the Hamiltonian~\cite{Adlong2020_Quasiparticle,Hu2023_Fermi}
\begin{align}
    \hat{H}_{\Omega}= \sum_{\kk}(\epsilon^I_{\kk} +\Delta_0) \dop_{\kk\dn}^\dagger \dop_{\kk\dn}+ \frac{\Omega_0}{2} \sum_\kk (\dop_{\kk \dn}^\dagger \dop_{\kk \up}+\text{h.c.}) \,.
\end{align}
The total Hamiltonian is then $\hat{H}_{\text{tot}} = \hat{H} + \hat{H}_{\Omega}$.

\subsection{Undriven impurity in a Fermi sea} \label{subsec:undriven}
Let us first recall the theories for the interacting impurity in the absence of Rabi drive ($\Omega_0=0$).

\subsubsection{Functional determinant approach}

In the case of an infinitely heavy impurity, it is possible to calculate the interacting impurity Green's function numerically exactly via the functional determinant approach (FDA).
In practice, the FDA allows one to calculate the Ramsey signal \cite{Schmidt2018_Universal,Drescher2024_Bosonic} as
 \begin{align} \label{eq:RamseyFDA}
    S(t)= \langle e^{i \hat{H}_0 t}e^{-i\hat{H} t} \rangle =\det \left[ \mathbb{1}-\hat{n}+\hat{n} e^{i\hat{h}_0t}e^{-i\hat{h}t}\right] \,,
\end{align}
where $\hat{H}_0$ and $\hat{H}$ correspond to the Hamiltonian without or with interactions between the infinitely heavy impurity and the medium, while  $\hat{h}_0,$ and $\hat{h}$ correspond to their single-particle representation, and $\hat{n}=(e^{\beta (\hat{h}_0-\mu\mathbb{1})} +\mathbb{1})^{-1}$ encodes the medium occupation with chemical potential $\mu$ and inverse temperature $\beta$. 

The above Ramsey signal is directly connected to the retarded Green's function as $G_{ \up}(t)=-i \theta(t) S(t)$.
Thus we can access the impurity Green's function in frequency space by taking the Fourier transform of the calculated $S(t)$,
\begin{align} \label{eq:GdnFDA}
G_{ \up}^\text{FDA}(\omega)=-i \int dt \theta(t) S(t)e^{i \omega t} \,.
\end{align}

\subsubsection{T-matrix approximation}
For an impurity of finite mass, there is no exact theory of the Fermi polaron. Nonetheless, the so-called $T$-matrix approximation has proven very successful to describe experiments near mass balance \cite{Massignan2026_Polarons}. In particular, this method is known to be equivalent to the variational Chevy ansatz \cite{Chevy2006_Universal} in the limit of zero temperature \cite{Combescot2007_Normal}. 
Within the $T$-matrix approximation, the zero-momentum impurity Green's function is 
\begin{align} \label{eq:GdnTmat}
G^T_{ \up}(\omega)=\frac{1}{\omega-\Sigma^T(\omega)} \,,
\end{align}
where the self-energy is given by 
\begin{align}\label{eq:selfE_ladderno_rabi}
    \Sigma^T(\omega)=\sum_\q n_\q  T(\q,\omega+\xi_{\q}) \,,
\end{align}
where $\xi_\q=\epsilon_{\q}-\mu$ and $n_\q=1/(\exp[\beta \xi_{\q}]+1)$ is the Fermi occupation of the medium. $T$ denotes the in-medium $T$ matrix,
\begin{align}\label{eq:Tmed}
T^{-1}(\q,\omega)= \frac{m_r}{2\pi a} -\sum_\kk  \left(\frac{1-n_\kk}{\omega-\xi_{\kk}-\epsilon^I_{\q-\kk }}+\frac{1}{\epsilon_\kk+\epsilon^I_\kk}\right) \,.
\end{align}

\subsubsection{Mass-gap model}
\label{massgap}
Recently, a new approach for finite-mass impurities has been introduced \cite{Chen2025_MassGap,Ramos2026_MassGapFDA}. It builds upon the FDA and includes finite mass effects via a gapped dispersion relation of the bath fermions in $\hat{H}_0$ and $\hat{H}$. At zero temperature, Eq.~\eqref{eq:RamseyFDA} and~\eqref{eq:GdnFDA} remain valid within the mass-gap model, with $\hat{H}_0$ and $\hat{H}$ replaced by their mass-gap counterparts. All expressions thus carry over unchanged at the level of the many-body determinants, and only the single-particle Hamiltonians entering the determinants are modified.

\subsection{Rabi-driven impurity in a Fermi sea} \label{subsec:driven_winter}

In presence of Rabi drive between the interacting $\ket{\up}$ impurity and the non-interacting $\ket{\dn}$ impurity, the zero-momentum impurity Green's function takes the general form \cite{Hu2023_Fermi,Mulkerin2024_Rabi}
\begin{align}\label{eq:G_inter}
\mathbf{G}(\omega)=
  \begin{pmatrix}
    \omega-\Sigma(\omega) & -\Omega_0/2\\[3pt]
    -\Omega_0/2 & \omega-\Delta_0
  \end{pmatrix}^{-1} \,,
\end{align}
where $\Sigma(\omega)$ is the medium-induced self-energy which accounts for the presence of interactions between medium fermions and the $\ket{\up}$ impurity. The above expression is formally exact and approximations arise only when evaluating the self-energy $\Sigma(\omega)$. In practice, we use different theories to evaluate Eq.~\eqref{eq:G_inter}. These theories are introduced in the following subsections.

\subsubsection{Coupled FDA} \label{subsec:FDA}

To model the system at small $\Omega_0$ for an infinitely heavy impurity ($M\rightarrow \infty$), we follow the method introduced in Ref. \cite{Adlong2021_Signatures}.
Specifically, it makes use of the fact that in the absence of Rabi coupling, we can calculate the interacting impurity Green's function numerically exactly via the FDA and then approximate the interacting Green's function matrix in presence of the Rabi coupling as
\begin{align} \label{eq:GFDA}
\mathbf{G}^{\text{FDA}}(\omega)&=\begin{pmatrix}
    G^{\text{FDA}}_{\up}(\omega)^{-1} &-\Omega_0/2 \\ -\Omega_0/2 &   \omega-\Delta_0
\end{pmatrix}^{-1}.
\end{align}
While $G^{\text{FDA}}_{ \up}(\omega)$ introduced in Eq.~\eqref{eq:GdnFDA} is exact in the absence of Rabi coupling, Eq.~\eqref{eq:GFDA} is approximate since it neglects the corrections of $ G^{\text{FDA}}_{\up}(\omega)$ due to the drive. Hence, we only expect Eq.~\eqref{eq:GFDA} to be accurate for small $\Omega_0/E_F$. We note, however, that this is precisely the regime where signatures of the AOC are expected to appear.

\subsubsection{Coupled T-matrix} \label{subsec:Tmatmodel1}

In the spirit of the previous Subsection, we can evaluate the self-energy in the $T$-matrix approximation in the absence of Rabi drive using Eq. \eqref{eq:selfE_ladderno_rabi}. Plugging it into \eqref{eq:G_inter} gives a similar approximation, as we introduced for the FDA approach, applied to the present $T$-matrix approach
\begin{align} \label{eq:GTmat_0}
\mathbf{G}^T(\omega)&=\begin{pmatrix}
    \omega-\Sigma^T(\omega) &-\Omega_0/2 \\ -\Omega_0/2 &   \omega-\Delta_0
\end{pmatrix}^{-1}.
\end{align}

\subsubsection{Full T-matrix} \label{subsec:Tmatmodel2}

Within the $T$-matrix approximation, the self-energy can also be evaluated including the Rabi drive \cite{Hu2023_Fermi,Mulkerin2024_Rabi} 
\begin{align}\label{eq:selfE_ladder_rabi}
    \Sigma^T_{\Omega}(\omega)=\sum_\q n_\q  T_\Omega(\q,\omega+\xi_{\q}) \,,
\end{align}
where $T_{\Omega}$ corresponds to the in-medium $T$ matrix \textit{in presence of the Rabi drive}, which reads \cite{Hu2023_Fermi,Mulkerin2024_Rabi,Bleu2025}
\begin{align}\label{eq:Tmed_Rabi}
&T_\Omega^{-1}(\q,\omega)= \frac{m_r}{2\pi a} %
-\sum_\kk  \left(\frac{c^2(1-n_\kk)}{\omega-\xi_{\kk}-\epsilon^I_{\q-\kk }-\epsilon_-}+\frac{s^2(1-n_\kk)}{\omega-\xi_{\kk}-\epsilon^I_{\q-\kk }-\epsilon_+} +\frac{1}{\epsilon_\kk+\epsilon^I_\kk}\right) \,.
\end{align}
Here, $\epsilon_{\pm}$ correspond to the dressed-impurity energies $\epsilon_{\pm}=\frac{1}{2}\left(\Delta_0\pm \sqrt{\Delta_0^2+\Omega_0^2}\right)$,
while $c^2,s^2$ are the fractions of the dressed states in the interacting $\ket{\up}$ state with the transformation coefficients satisfying $c^2=\frac{1}{2}(1+\frac{\Delta_0^2}{\sqrt{\Delta_0^2+\Omega_0^2}}) $, $cs=\frac{\Omega_0}{2\sqrt{\Delta_0^2+\Omega_0^2}}$ and $c^2+s^2=1$. 
Crucially, Eqs.~\eqref{eq:selfE_ladder_rabi} and \eqref{eq:Tmed_Rabi} account for the possibility of the impurity to Rabi flip between collisions. The resulting theory is non-perturbative in $\Omega_0$ and it is thus accurate for large $\Omega_0/E_F$. 
The Green's function matrix in this case becomes
\begin{align} \label{eq:GTmat_full}
\mathbf{G}^T_\Omega(\omega)&=\begin{pmatrix}
   \omega-\Sigma^T_\Omega(\omega)  &-\Omega_0/2 \\ -\Omega_0/2 &    \omega-\Delta_0
\end{pmatrix}^{-1}.
\end{align}

\section{Extracting polaron properties from simulations} 

\subsection{Polaron properties without Rabi drive} \label{sec:Gamma_Z}

When the polaron quasiparticle is well-defined with an energy $E_\mathrm{p}$, the impurity Green's function can be approximated as 
\begin{align} \label{eq:G_norabi_res}
   G_{\up}(\omega) \simeq \frac{Z}{\omega-E_\mathrm{p}+i\Gamma_0} \,,
\end{align}
in the vicinity of $\omega=E_\mathrm{p}$ \cite{Massignan2014_Polarons}.
Here, $Z$ and $\Gamma_0$ are the polaron quasiparticle residue and width that can be obtained from the impurity self-energy $\Sigma( \omega)$ as
\begin{equation} \label{eq:Z_and_Gamma}
	Z = \left[1-\left.\frac{\partial \,\mathrm{Re} \Sigma( \omega)}{\partial \omega}\right|_{\omega=E_\mathrm{p}}\right]^{-1},\quad \Gamma_0 = -Z \, \mathrm{Im} \Sigma( E_\mathrm{p}).
\end{equation}

We note that at zero temperature, the Lorentzian form in Eq.~\eqref{eq:G_norabi_res} is incorrect for the infinitely heavy impurity because of AOC (see Sec.~\ref{sec:AOC_prefactors}). Nonetheless, such a Lorentzian shape emerges at finite temperatures since the Ramsey signal acquires a thermal exponential decay tail at long times $|S(t)| \simeq Z \exp(-\Gamma t)$ \cite{Schmidt2018_Universal}. Thus, one can also obtain some peak weight $Z$ and width $\Gamma$ from FDA calculations at finite temperatures.

\subsection{Polaron properties with Rabi drive}
\label{subsec:extraction}

\paragraph{Using pole fitting ---} \label{par:pole_fitting}
To evaluate the renormalized Rabi frequencies and the oscillation damping rates theoretically, we use the fact that we can approximate the Green's function by keeping only the contribution of the two quasi-particle peaks at $\omega=E_{\pm}$,
\begin{align} \label{eq:pole_fitting}
    \mathbf{G}_{22}(\omega) \simeq \frac{Z_+}{\omega - E_+ + i\Gamma_+} + \frac{Z_-}{\omega - E_- + i\Gamma_-} \,,
\end{align}
where $E_{\pm},\Gamma_{\pm}$ and $Z_{\pm}$ denote the peak energies, widths and weights. 
In general, we find that the lower peak  $E_-$ has a small broadening $\Gamma_-$, while the higher-energy pole can have substantial broadening $\Gamma_+>\Gamma_{-}$.  

In the time-domain, \eqref{eq:pole_fitting} leads to damped oscillations with a renormalized Rabi frequency and damping given by
\begin{align}\label{eq:split_inter2}
\Omega = E_+ - E_-\,, \quad
\Gamma=\Gamma_++\Gamma_- \,.
\end{align}
In order to obtain all the theoretical curves presented in the main text, we calculate \eqref{eq:G_inter} numerically using the different theories introduced above and extract $\Omega$ and $\Gamma$ by fitting the calculated $\mathbf{G}_{22}(\omega)$ with Eq.~\eqref{eq:pole_fitting}. We find that this method works very well to analyze the driven attractive polaron on the negative side of the resonance $a<0$, and we also use it to compare the different theories in regimes not accessed in our experiments in Figures \ref{fig:TTF_scan} and \ref{fig:mass_ratio_scan}.

\paragraph{Using pseudo-properties ---} \label{par:pseudo_Z_Gamma}
On the other hand, we find that the above approach for extracting $\Omega$ and $\Gamma$ becomes unreliable to investigate the driven repulsive polaron on the other side of the resonance ($a>0$). This is because the corresponding spectral function can now exhibit three relevant peaks due to the presence of the attractive polaron, rendering the approximation given by Eq.~\eqref{eq:pole_fitting} inaccurate. 
While we could, in principle, use a three-poles approximation, such an approach would introduce multiple oscillation frequencies and damping rates that cannot be easily interpreted and compared with the single ones obtained in the experiments. 
Thus, to compare our experimental results with calculations in this regime, we employ the method used in Ref.~\cite{Vivanco2025_strongly}, which relies on pseudo-properties of polarons.
Specifically, one defines a pseudo-residue $\overline{Z}$ and a pseudo-width $\overline{\Gamma}$ via Eq.~\eqref{eq:Z_and_Gamma} evaluated at the undriven polaron energy $E_\text{p}$, but with the self-energy calculated in the presence of the external drive using Eq.~\eqref{eq:selfE_ladder_rabi}. The results of this method are compared with experimental data in Fig.~\ref{fig:Rabi_vs_drive_SM}(c-e), where we plot $\overline{\Gamma}$ (top panel) and $\overline{Z} + (\overline{\Gamma}/\Omega_0)^2$ (bottom panel) as dotted black lines, analogous to $(\Omega/\Omega_0)^2 = Z + (\Gamma/\Omega_0)^2$ (see main text).

\newpage

\section{AOC-scaling prefactors at $T=0$} \label{sec:AOC_prefactors}
In this Section, we explain how we can obtain theoretically the AOC prefactors at $T=0$.
To do so, we make use of the fact that for an infinitely heavy impurity, the Ramsey signal $S(t)$ can be calculated numerically exactly via the FDA \cite{Schmidt2018_Universal} and that there exist analytical results for its long time asymptotic behavior at $T=0$ \cite{Nozieres1969_Singularities,Combescot1971,Knap2012_TimeDependent}
  \begin{align}\label{eq:Stail_T0}
    S(t)\simeq \frac{C e^{-i \Delta E t} }{(i t/t_F)^\alpha}+ \theta(a) \frac{C_b e^{-i (\Delta E-E_F+E_B) t}}{(i t/ t_F)^{\alpha_b}} \,.
\end{align}
Here, $C$ and $C_b$ are dimensionless prefactors which depend on the interaction parameter $1/(k_Fa)$, and
\begin{align}
\alpha=\left(\frac{\delta_F}{\pi}\right)^2\,, \quad
\alpha_b=\left(1+\frac{\delta_F}{\pi}\right)^2\,, \quad
 \delta_F=  -\arctan(k_F a)\,, \quad
 \Delta E=-\int_{0}^{E_F}\frac{dE}{\pi} \delta(\sqrt{2mE})\,.
\end{align}
The second term in Eq.~\eqref{eq:Stail_T0} is only present on the repulsive side of the resonance when $a>0$, where a bound-state exists and $E_B=-\hbar^2/(2ma^2)$. Physically, Eq.~\eqref{eq:Stail_T0} encodes the essence of the AOC, namely, that the overlap $|S(t)|\rightarrow 0$ as a power law for $t\rightarrow \infty$.
\begin{figure}[b!] 
    \includegraphics[width=0.5\linewidth]{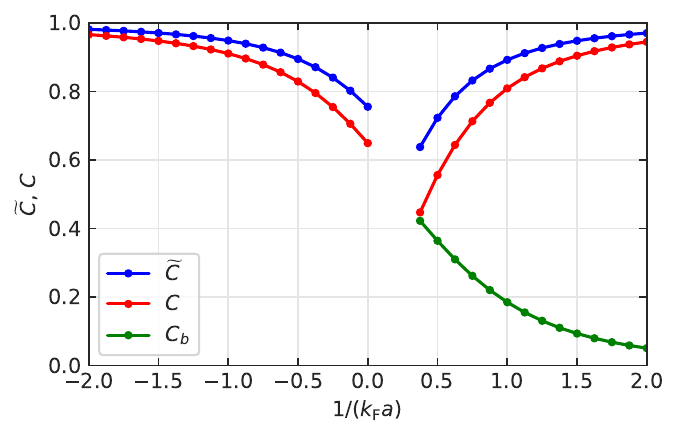}
    \caption{Prefactors $C$, $C_b$ of the long time tail of $S(t)$, Eq.~\eqref{eq:Stail_T0} versus $1/(k_\mathrm{F} a)$ together with the corresponding prefactor $\widetilde{C}$ appearing in the renormalized Rabi splitting Eq.~\eqref{eq:Rabipowerlaw}. }
    \label{fig:prefactor_AOC}
\end{figure}

As discussed in the main text, one expects that the response of the impurity to a Rabi drive will inherit signatures of the orthogonality catastrophe.
Specifically, by analyzing the poles of the zero-temperature Green's function for small $\Omega_0/E_F$, it was shown that the renormalized Rabi splitting follows a power law \cite{Adlong2021_Signatures}
\begin{align}\label{eq:Rabipowerlaw}
    \frac{\Omega}{E_F}= \widetilde{C}  \left(\frac{\Omega_0}{E_F}\right)^{\frac{2}{2-\alpha}} \,,
\end{align}
where $\widetilde{C}$ is a $1/(k_F a)$-dependent prefactor related to $C$ in the asymptotic Ramsey signal above as
\begin{align}
   \widetilde{C} = \left(\frac{4}{ \Gamma[1-\alpha] C}\right)^{\frac{1}{\alpha-2}} \left(1+ \cos\left[\frac{\pi \alpha}{\alpha-2}\right] \right) \,,
\end{align}
where $\Gamma[x]$ is the Gamma-function. From Eq.~\eqref{eq:Rabipowerlaw}, we have $\Omega/\Omega_0\propto (\Omega_0/E_F)^{\alpha/(2-\alpha)}\rightarrow 0$ as a power law when $\Omega_0/E_F\rightarrow 0$, a behavior which differs from the case where a well-defined quasiparticle is Rabi-driven, which would approach $\Omega/\Omega_0\simeq \sqrt{Z}$ in this limit \cite{Kohstall2012_Metastability}.

While we are not aware of explicit analytical expressions giving the dependence of $C$ in terms of the interaction parameter, it can be accessed numerically \cite{Knap2012_TimeDependent} by fitting the long time tail of the $|S(t)|$ calculated using FDA with the analytical expression \eqref{eq:Stail_T0}.
The results of this procedure are presented in Figure \ref{fig:prefactor_AOC}, where we have plotted the coefficients $C$ and $\widetilde{C}$ obtained from the fits.  For completeness, we also include the fitted $C_b$ on the positive side even though it does not appear in the renormalized Rabi scaling law \eqref{eq:Rabipowerlaw}.

\section{Detailed experimental data and theory comparison}

In this Section, we provide additional experimental and theoretical data supporting the results in the main text. 
Sec.~\ref{subsec:Rabi_resp} presents detailed experimental data of the driven polaron for different scattering lengths including fits to extract the exponents of the observed scaling law and comparisons with theory.
In Sec.~\ref{subsec:Emergence_theo}, we use the coupled FDA and full $T$-matrix theory to study the emergence of the scaling law as a function of temperature and impurity-bath mass ratio. 
In \ref{subsec:Rabi_vs_inter}, we compare the measured Rabi frequencies with the complete $T=0$ AOC predictions and with quasiparticle weights $Z$ evaluated within different theories.
In Sec.~\ref{subsec:Rabi_vs_det}, we discuss the overshoots of $\Omega/\Omega_0>1$ at strong drive by showing Rabi frequencies as a function of detuning.
Finally, Sec. \ref{subsec:NIBA} presents a comparison between experimental data and non-interacting blip approximation (NIBA) predictions for $\Omega/\Gamma$.

\subsection{Rabi response of the driven polaron as a function of interactions} \label{subsec:Rabi_resp}

In this Subsection, we present the experimentally measured Rabi frequencies and damping rates in detail for all impurity-bath interactions $1/(k_\mathrm{F}a)$ probed in our experiments.

\begin{figure}
    \centering
    \includegraphics[width=0.95\linewidth]{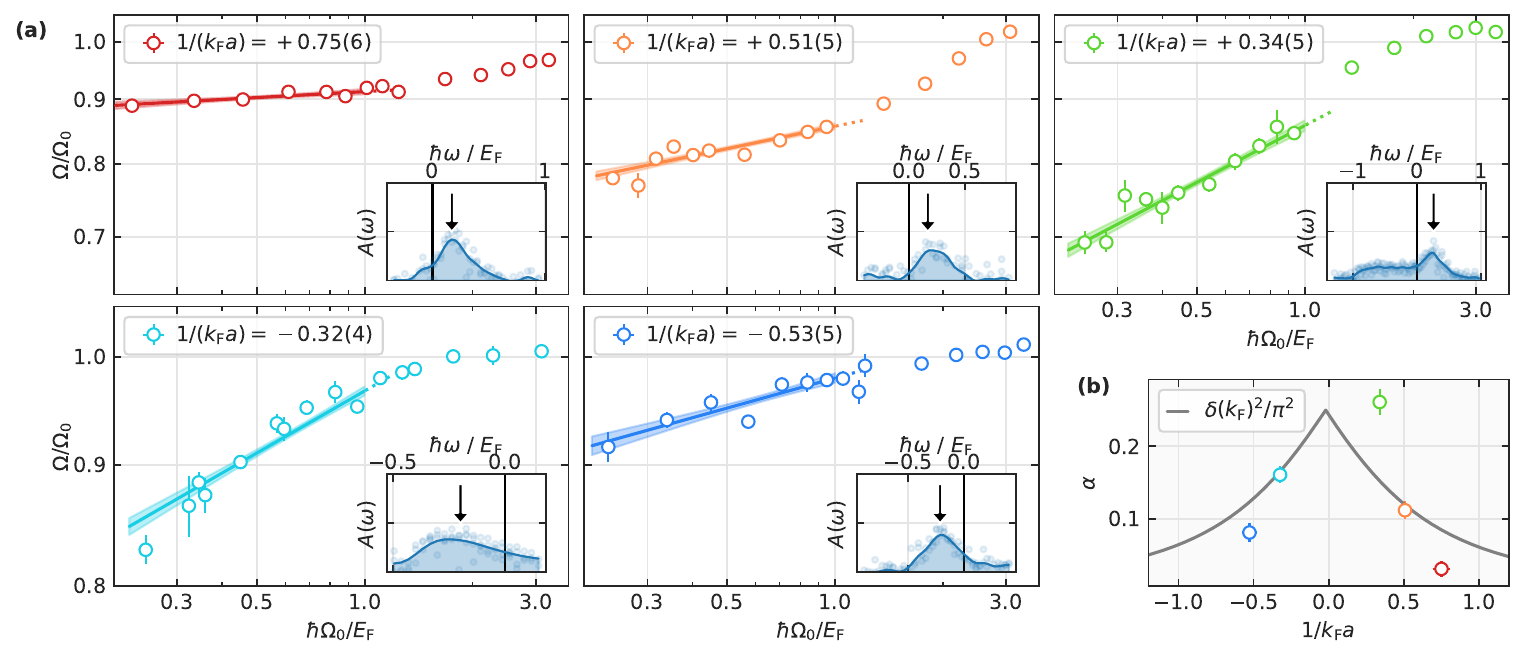}
    \caption{Extended Fig.~2 of the main text, including the fits for all five interaction parameters \textbf{(a)} for the repulsive (top row) and attractive polaron (bottom row). The injection spectra for each impurity-bath interaction are shown as insets, and the arrows mark the energy of the drive used for the corresponding measurement. The fitted exponents are shown in \textbf{(b)} with matching colors.}
    \label{fig:Rabi_AOCfits_all}
\end{figure}

In Figure~\ref{fig:Rabi_AOCfits_all}, we show the different fits of $\Omega/\Omega_0$ with the AOC scaling law [Eq.~(1) of the main text] for different scattering lengths. Here, we use both the prefactors and the scaling exponents $\alpha$ as fit parameters. This fitting procedure was used to obtain the experimental scaling exponents shown in Fig.~2 of the main text, which we have reproduced here in panel (b) with the matching color scale for clarity.

\begin{figure}[!ht]
    \centering
    \includegraphics[width=0.8\linewidth]{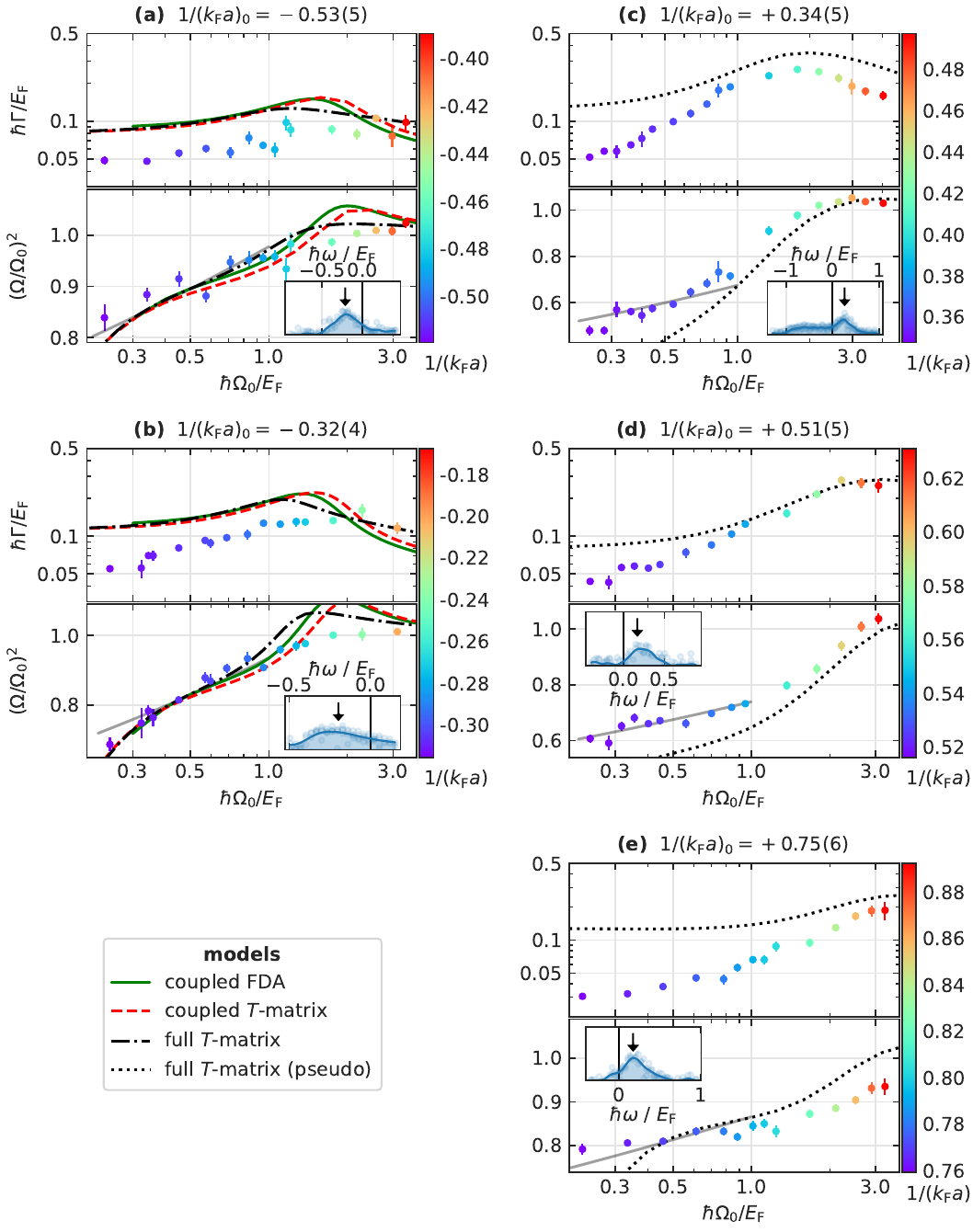}
    \caption{Detailed comparison of the response of the driven {attractive} \textbf{(a)}-\textbf{(b)} and repulsive \textbf{(c)}-\textbf{(e)} polaron as a function of drive strength (detuning of the Rabi drive is indicated in the insets). For each interaction strength we show the experimentally measured damping rates (upper panel) and $(\Omega/\Omega_0)^2$ (lower panel) together with the predictions from theories, described in Sec.~\ref{sec:model}. For the attractive polaron (a,b) the calculated $\Gamma$ and $\Omega$ are extracted from pole fitting of the three theories with Eq.~\ref{eq:pole_fitting}. The dotted line shown for the repulsive polaron (c-e) are based on the full $T$-matrix theory but using the pseudo-properties $\bar{Z}$ and $\bar{\Gamma}$ (\textit{cf.} Sec.~\ref{subsec:extraction}) instead of pole fitting. The scaling prediction of Eq.~(1) of the main text is shown as solid gray line where the prefactor has been adapted to match the data. The colors indicate the changing interaction parameter $1/(k_\mathrm{F}a)$ as a function of drive strength, see Sec.~\ref{sec:FR_shift}. }
    \label{fig:Rabi_vs_drive_SM}
\end{figure}

In Figure~\ref{fig:Rabi_vs_drive_SM}, we plot the experimental results for both $\Gamma$ and $(\Omega/\Omega_0)^2$ keeping track of the small changes in $1/(k_F a)$ indicated by the color scale (cf. Sec. \ref{sec:FR_shift}). 
For the attractive polaron [panel (a) and (b)], we compare the experimental data with the different theories as in Fig.~3 of the main text, where $\Gamma$ and $\Omega$ are determined by fitting Eq.~\eqref{eq:pole_fitting} to the resulting Green's functions, as described in Sec.~\ref{subsec:extraction}. 
We find a good quantitative agreement between the measured $(\Omega/\Omega_0)^2$ and the ones calculated with the different theories for $\Omega_0\lesssim E_F$. For $\Omega_0\gtrsim E_F$, however, the results obtained from the full $T$-matrix theory are the most accurate, since it incorporates the Rabi coupling to all orders in the self-energy. 
As mentioned in the main text, while the non-monotonic behavior of the calculated $\Gamma$ is in qualitative agreement with the experimentally measured one, we observe that the theory overestimates the magnitude of the damping rates for $\Omega_0\lesssim E_F$.
In panels (c-e), we show the results for the driven repulsive polaron. As explained in Sec.~ \ref{subsec:extraction}, in this case it is challenging to extract $\Omega$ and $\Gamma$  unambiguously using the same method. Nonetheless, inspired by the the comparison made in Ref.~\cite{Vivanco2025_strongly}, we can compare the experimental results with the calculated pseudo-polaron properties (dotted-black line). Similarly to Ref.~\cite{Vivanco2025_strongly}, we find a reasonable agreement with the experimental data.
Finally, in all panels showing $(\Omega/\Omega_0)^2$, we also plot the AOC scaling predictions as solid gray lines up to drive strengths of $\hbar \Omega_0 \leq E_\mathrm{F}$. Note that, here, in contrast to Fig~\ref{fig:Rabi_AOCfits_all}, the scaling exponent is not a fit parameter, rather we use $\alpha=\delta(k_\mathrm{F})^2/\pi^2$ calculated from the experimental scattering length. Overall, we observe a good agreement between the AOC scaling and the experimental observations.

\subsection{Theory comparison and emergence of the Anderson orthogonality scaling law } \label{subsec:Emergence_theo}

Here we compare coupled FDA and full $T$-matrix predictions for the Rabi-driven polaron to see how the scaling law emerges as a function of temperature and impurity-bath mass ratio.
For these comparisons, we focus on the case of the attractive polaron and fix the interaction parameter to $1/(k_\mathrm{F}a) = -0.32$ as in Fig.~3 of the main text.

\begin{figure}
    \centering
    \includegraphics[width=0.9\linewidth]{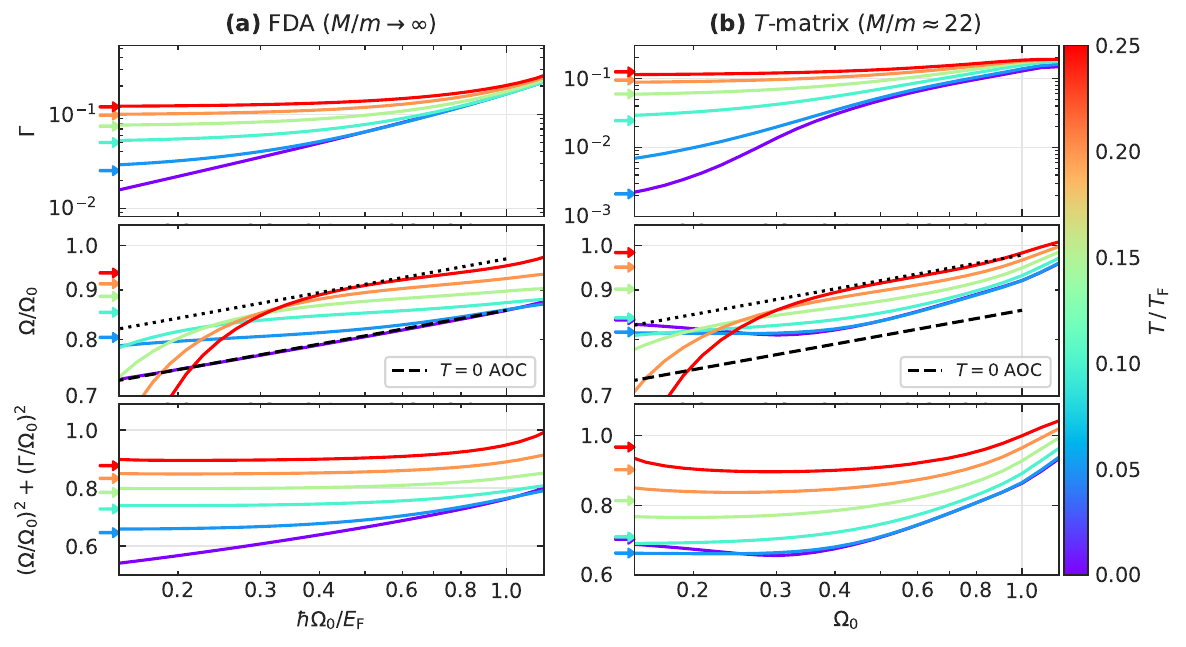}
    \caption{
    Theory predictions at $1/(k_\mathrm{F}a) = -0.32$ for increasing temperatures $T/T_\mathrm{F} = 0,\, 0.05,\, 0.1,\, 0.15,\, 0.2,\, 0.25$. \textbf{(a)} Coupled FDA results ($M/m \to \infty$), 
    \textbf{(b)} full $T$-matrix results with $M/m=22$. 
    In both panels, $\Gamma$ and $\Omega$ are extracted by fitting the calculated Green's function with Eq.~\eqref{eq:pole_fitting}. 
    The colored arrows indicate polaron properties in the absence of external drive: $\Gamma_0$ (top), $\sqrt{Z}$ (middle), and $Z$ (lower panel), calculated using FDA (a) and $T$-matrix (b). The dashed-black lines show the AOC scaling predictions with the calculated $T=0$ prefactor (\textit{cf.} Sec.~\ref{sec:AOC_prefactors}), while the dotted-black lines show the AOC power law with an arbitrary prefactor. 
    }
    \label{fig:TTF_scan}
\end{figure}

\paragraph{Temperature dependence ---}
In Fig.~\ref{fig:TTF_scan}, we show the results of both theories for increasing temperatures $T/T_\mathrm{F}$ (colored solid lines). We find that the coupled FDA calculations (Fig.~\ref{fig:TTF_scan}a), valid for an infinitely heavy impurity, recover the expected $T=0$ AOC scaling law for $\Omega/\Omega_0$ in Eq.~\eqref{eq:Rabipowerlaw}, including the correct prefactor (Sec.~\ref{sec:AOC_prefactors}). Already increasing the temperature to $T/T_\mathrm{F} = 0.05$ results in a  significant quasiparticle residue $Z$, as indicated by the arrows on the left of the second and third panel. At this low temperature, the damping rate $\Gamma$ is still small (upper panel where the arrows indicate the polaron spectral width $\Gamma_0$ in the absence of a drive). However, at higher temperatures, the damping becomes significant so that the $(\Omega/\Omega_0)^2 \approx Z$ plateau for small $\Omega_0$ can only be observed after compensating for the damping effect (lower panel). At $T>0$, the nonzero quasiparticle residue restricts the range of drive strengths where $\Omega/\Omega_0$ follows a power law, also weakly affecting the observed exponent. %
Small deviations from the $T=0$ exponent are expected at finite temperature, as reported in \cite{Adlong2021_Signatures}, however, we find slightly different numerical results at the lowest temperatures.
At the experimental temperature, the predictions recover a scaling exponent similar to the $T=0$ result (cf. black dotted line in Fig.~\ref{fig:TTF_scan}a).  Overall, these finite-temperature deviations are small compared to the interaction dependence of $\alpha$ at a fixed temperature, \textit{cf.} Fig.~\ref{fig:Rabi_AOCfits_all}.

The full $T$-matrix (Fig.~\ref{fig:TTF_scan}b), which retains only single particle-hole excitations of the Fermi sea, does \emph{not} reproduce the AOC scaling at low temperature. Instead, already at $T=0$, it converges to a finite quasiparticle weight $Z$. Note that the non-monotonous behavior of the quasiparticle properties at small temperatures was also observed in \cite{Hu2022_Fermi}. 
However, as the temperature increases, the two theories become increasingly similar. At the highest temperatures, also the full $T$-matrix calculations show a similar power-law scaling (dotted lines, \textit{cf.} Fig.~3 of the main text).

\begin{figure}
    \centering
    \includegraphics[width=0.9\linewidth]{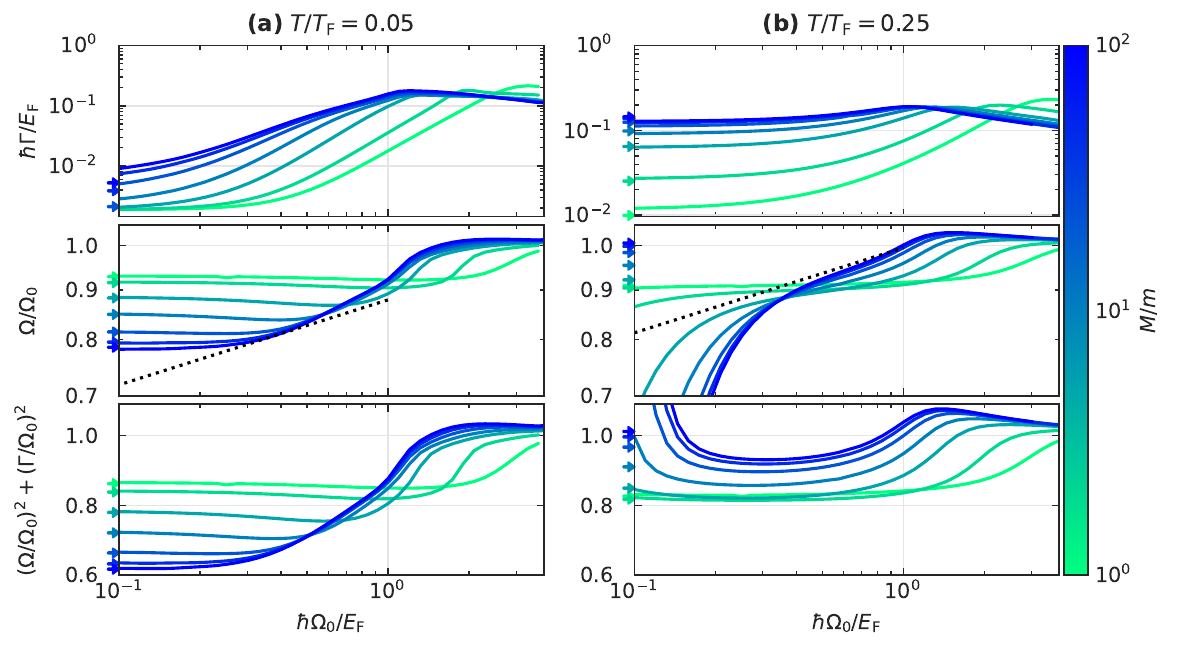}
    \caption{Full $T$-matrix predictions, where $\Gamma$ and $\Omega$ are extracted from fitting Eq.~\eqref{eq:pole_fitting},  for varying impurity-bath mass ratios $M/m$ (color coded) at fixed temperature \textbf{(a)} $T=0.05\,T_\mathrm{F}$ and \textbf{(b)}  $T=0.25\,T_\mathrm{F}$ and interaction parameter $1/(k_\mathrm{F}a) = -0.32$. The mass ratios are $M/m =$ 1, 2, 5, 10, 22, 46, and 100.
    The black dotted line is the AOC scaling prediction (with arbitrary prefactor). The divergence at small drive strengths in the lowest panel in (b) signals the approach to the regime of overdamped Rabi oscillations. Arrows as in Fig.~\ref{fig:TTF_scan}b.}
    \label{fig:mass_ratio_scan}
\end{figure}

\paragraph{Mass-ratio dependence ---}
To investigate how the $T$-matrix calculations converge to a power-law behavior at high temperatures and large impurity-bath mass ratios $M/m$, Fig.~\ref{fig:mass_ratio_scan} shows full $T$-matrix simulations at two different temperatures as a function of increasing $M/m$. 

At fixed interaction, the polaron spectrum and quasiparticle properties change considerably with $M/m$ (see the evolving $\Gamma$ and $Z$, as indicated by the arrows). 
For both temperatures, the power law builds up with increasing mass ratio and is convincingly recovered for $M/m \gtrsim 10$, setting in at strong drive $\hbar\Omega_0 \approx E_\mathrm{F}$ and extending to weaker drive before being cut off by the finite impurity mass.

At the lowest temperature $T/T_\mathrm{F}=0.05$, the recovered scaling is not fully consistent with the $T=0$ AOC prediction. We attribute this residual discrepancy to finite temperature rather than to the single particle-hole truncation. In the regime $Z\approx 1$, the single particle-hole model is well justified. We have verified independently, using the mass-gap theory applied to the Rabi-driven system, that the correct scaling is recovered at exactly $T=0$ as $M/m$ increases.

\paragraph{Implications for the experiment.}
Together, these results show that the Cs-Li mass ratio $M/m \approx 22$ is already sufficient to observe power-law scaling in $\Omega/\Omega_0$ over the range of drive strengths probed in the main text. This numerical study also reveals that for heavier impurities the expected damping rates are generally larger. This result is expected from the spectral width in the absence of an external drive at $T>0$. %

\begin{figure}
    \centering
    \includegraphics[width=0.4\linewidth]{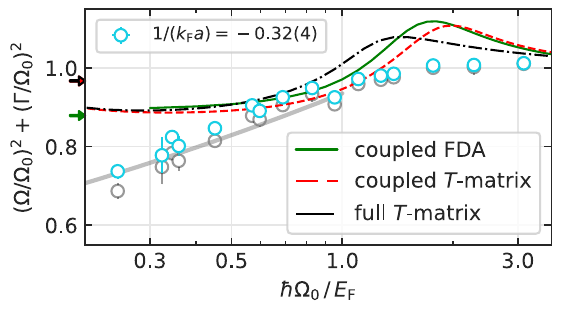}
    \caption{Same data as in Fig.~3a of the main text, now accounting for the experimentally measured damping $\Gamma$ via $(\Omega/\Omega_0)^2 + (\Gamma/\Omega_0)^2$ (blue markers). The usual $(\Omega/\Omega_0)^2$ without additional modifications is shown as gray open markers. As in Fig.~3a, the gray solid line shows the $T=0$ AOC prediction. The different model predictions are also calculated according to $(\Omega/\Omega_0)^2 + (\Gamma/\Omega_0)^2$, where $\Omega/\Omega_0$ and $\Gamma$ are shown in Fig.~3 of the main text.}
    \label{fig:damping_correction_experiment}
\end{figure}

We note that the effect of the large damping rates on the reduction of Rabi frequencies described for both FDA and $T$-matrix predictions is not limiting our experimental extraction of $\Omega/\Omega_0$, due to much smaller measured values of $\Gamma$. This is illustrated in Fig.~\ref{fig:damping_correction_experiment}, where we compare the measured $(\Omega/\Omega_0)^2$ with and without accounting for the damping via $(\Omega/\Omega_0)^2 + (\Gamma/\Omega_0)^2$, that we used for the theory plots above.

\subsection{Renormalized Rabi frequency as a function of interactions} \label{subsec:Rabi_vs_inter}

\begin{figure}
    \centering
    \includegraphics[width=0.9\linewidth]{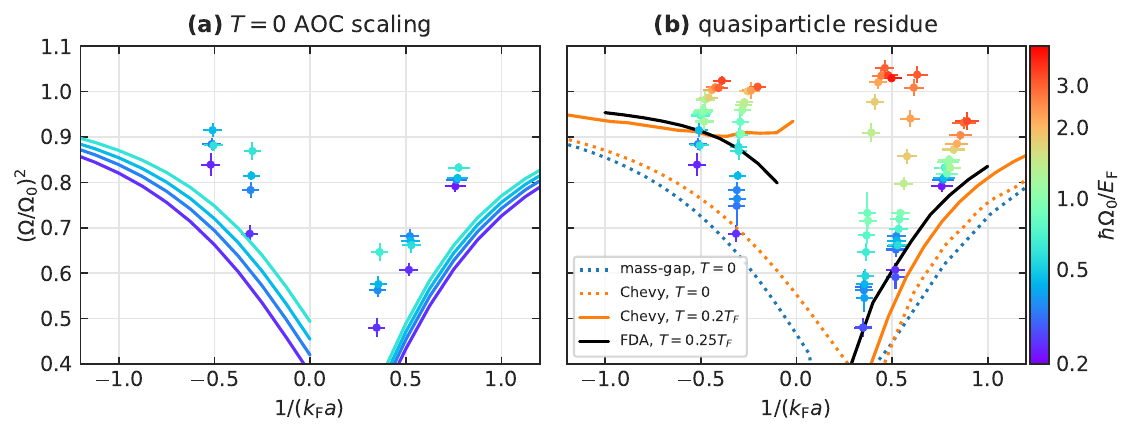}
    \caption{Reduction of the driven heavy Fermi polaron Rabi frequency as a function of drive strength $\hbar\Omega_0$ (color coded) and impurity-bath interactions $1/(k_ \mathrm{F}a)$. Both panels show the same experimental data as presented in Fig.~\ref{fig:Rabi_vs_drive_SM}. Vertical error bars indicate the fit uncertainties while the horizontal error bars represent systematic uncertainties on the interaction parameters, dominated by uncertainties on $k_\mathrm{F}$. 
    In \textbf{(a)}, the solid lines show the AOC power-law predictions of Eq.~\eqref{eq:Rabipowerlaw} with the $T=0$ prefactors (omitted in the main text, \textit{cf.} Sec.~\ref{sec:AOC_prefactors}).
    In \textbf{(b)}, we show the experimental data along theory predictions for the quasiparticle weight  $Z$ (Eq.~\eqref{eq:Z_and_Gamma}) calculated within different models, detailed in Sec.~\ref{subsec:undriven}, either at $T=0$ (dotted lines) or close to the experimental temperature (solid lines).}
    \label{fig:Rabi_vs_interactions}
\end{figure}

In this Subsection, we take a closer look at the dependence of the measured ratio $(\Omega/\Omega_0)^2$ on the impurity-bath interactions.
In Figure.~\ref{fig:Rabi_vs_interactions}, the colored filled circles correspond to the experimental measurements where the color scale encodes the strength of the Rabi drive $\Omega_0$.
In panel (a), we compare the experimental data for low drive strengths with the AOC predictions at the corresponding $\Omega_0$ shown by colored solid lines. To plot the AOC lines, we use the analytical predictions of Eq.~\eqref{eq:Rabipowerlaw} with the calculated prefactors (omitted in Eq.~(1) of the main text) as explained in detail in Sec.~\ref{sec:AOC_prefactors}. 
Although we do not expect our finite temperature experimental measurement to perfectly reproduce the zero temperature AOC results, we find good qualitative agreement in the dependence of $\Omega/\Omega_0$ as a function of interactions as well as on the trend versus $\Omega_0$. We can also observe that the AOC scaling captures the asymmetry between the attractive and repulsive side of the resonance which can be seen in the experimental data.

 As explained in the introduction of the main text, in contrast to the AOC scaling, for a mobile impurity the Chevy ansatz theory at zero temperature predicts $(\Omega/\Omega_0)^2\rightarrow Z$ as $\Omega_0\rightarrow 0$.
For completeness, in Fig.~\ref{fig:Rabi_vs_interactions}(b), we compare the same experimental data with the quasiparticle residue $Z$ predicted by different theories: FDA and $T$-matrix approximation at finite temperature, as well as the $T$-matrix approximation and mass-gap model at zero temperature (cf. Sec. \ref{subsec:undriven}). 
We can see that the predictions of the FDA, the zero-temperature $T$-matrix, and the mass-gap model all capture qualitatively the observed trend as a function of $1/(k_\mathrm{F}a)$, although with different magnitudes. 
Interestingly, the zero temperature $T$-matrix predictions (labeled ``Chevy"), seem to capture the qualitative dependence better then their finite temperature counterparts, consistent with earlier observations \cite{Kohstall2012_Metastability, Scazza2017_Repulsive}. We also note that at $T=0$, the $T$-matrix predictions are close to those of the mass-gap model (Sec.~\ref{massgap}) which is expected to be more accurate for very large mass ratios \cite{Chen2025_MassGap}.

\subsection{Rabi frequency as a function of detuning} \label{subsec:Rabi_vs_det}  %

\begin{figure}
    \centering
    \includegraphics[width=0.5\linewidth]{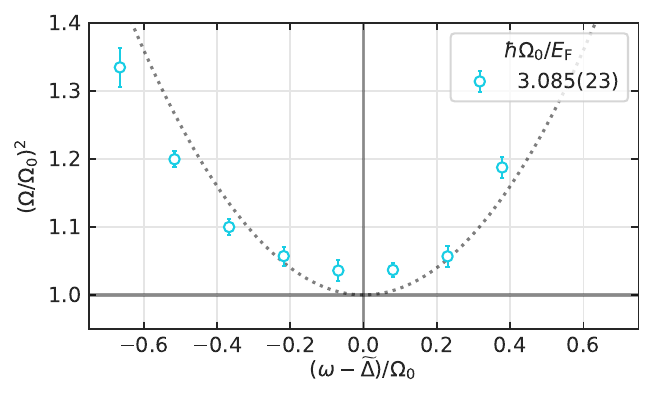}  %
    \caption{Interacting Rabi frequency $(\Omega/\Omega_0)^2$ as a function of detuning $(\omega-\widetilde{\Delta}) / \Omega_0$ for $1/(k_\mathrm{F}a) = \num{-0.32(4)}$ and strong drive $\hbar\Omega_0\approx3E_\mathrm{F}$. Here $\widetilde{\Delta}$ is the energy of the driven polaron determined as the zero-crossing of the steady-state magnetization $\widetilde{\mathcal{M}}(\omega=\widetilde{\Delta})=0$, a concept established in \cite{Vivanco2025_strongly}.
    The dotted black line shows the two-level limit $(\Omega/\Omega_0)^2 = 1 + (\omega-\widetilde{\Delta})^2/\Omega_0^2$.}
    \label{fig:Rabi_detuing_OmRed}
\end{figure}

For drive strengths around $\hbar \Omega_0 \approx 3 E_\mathrm{F}$ we observe interacting Rabi frequencies exceeding the bare Rabi frequencies. To exclude that this surprising ``overshoot" is a measurement artefact caused by a finite detuning from the polaron energy, we vary the detuning of the Rabi coupling around the polaron energy, see Fig.~\ref{fig:Rabi_detuing_OmRed} and extract the resulting change in Rabi frequency reduction. For all detunings, we observe interacting Rabi frequencies $\Omega > \Omega_0$. 

\subsection{Zero-temperature relation of Rabi frequency and damping} \label{subsec:NIBA}
Finally, we can compare our results with the predictions of the spin-boson model of a Rabi-driven static impurity \cite{Leggett1987_Dynamics}. At zero temperature and in the non-interacting blip approximation (NIBA), the damping and frequency of the Rabi oscillations are predicted to be universally related by \cite{Leggett1987_Dynamics, Knap2013_Dissipative}
\begin{equation} \label{eq:Omega_over_Gamma_NIBA}
    \Omega/\Gamma = -\tan \frac{\pi}{2-\alpha} \,.
\end{equation}
Although this relation is not expected to hold at finite temperatures, and longer interrogation times, one might still look for a regime of roughly constant $\Omega/\Gamma$, when $\hbar\Omega_0 < E_\mathrm{F}$ and $\Gamma$ is neither dominated by thermal broadening nor by the repulsive polaron lifetime.

\begin{figure}
    \centering
    \includegraphics[width=0.7\linewidth]{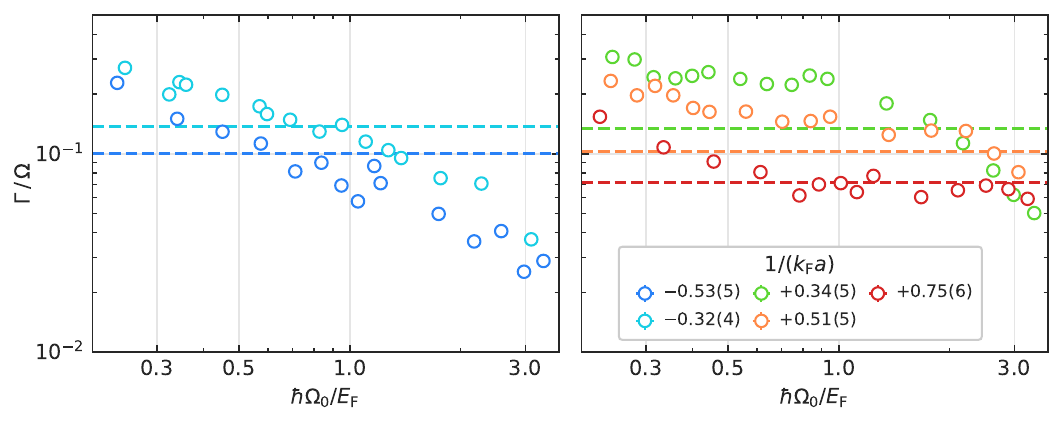}
    \caption{Ratio of experimentally determined damping rate $\Gamma$ and frequency $\Omega$ of the Rabi oscillations for the attractive (left) and repulsive polaron (right panel). The constant $T=0$ NIBA \cite{Leggett1987_Dynamics} prediction of Eq.~\eqref{eq:Omega_over_Gamma_NIBA} are shown as dashed lines in matching colors.}
    \label{fig:Gamma_over_Omega}
\end{figure}
We test the relation of $\Omega$ and $\Gamma$ by showing the ratio $\Gamma/\Omega$ in Fig.~\ref{fig:Gamma_over_Omega} together with the prediction of Eq.~\eqref{eq:Omega_over_Gamma_NIBA}.
For the attractive polaron (left panel), we do not find extended regimes of constant $\Gamma/\Omega$, still, in the range of $0.5\lesssim \hbar\Omega_0/E_\mathrm{F} \lesssim 1$, \textit{i.e.} also where the AOC scaling law describes the measured $\Omega/\Omega_0$ well, quantitative deviations of the measured $\Gamma/\Omega$ from the NIBA prediction are small. On the repulsive side (right panel) the quantitative agreement is slightly worse, whereas the measured $\Gamma/\Omega$ show a clearer plateau in the regime of drive strengths mentioned above. 
Overall, although Eq.~\eqref{eq:Omega_over_Gamma_NIBA} is a zero-temperature, short-time approximation, its predictions are broadly consistent with the measured data.

\FloatBarrier
\bibliography{main}%